\documentclass[aps,letterpaper,floatfix,twocolumn]{revtex4-1}
\usepackage{graphicx,amssymb,amsmath}

\begin{document}

\title{Directly measuring single molecule heterogeneity using force
  spectroscopy}

\author{Michael Hinczewski}
\affiliation{Department of Physics, Case Western Reserve University, OH 44106}

\author{Changbong Hyeon}
\affiliation{Korea Institute for Advanced Study, Seoul 130-722, Korea}

\author{D. Thirumalai}
\affiliation{Department of Chemistry, The University of Texas at Austin, TX 78712}

\begin{abstract}
  One of the most intriguing results of single molecule experiments on
  proteins and nucleic acids is the discovery of functional
  heterogeneity: the observation that complex cellular machines
  exhibit multiple, biologically active conformations. The structural
  differences between these conformations may be subtle, but each
  distinct state can be remarkably long-lived, with random
  interconversions between states occurring only at macroscopic
  timescales, fractions of a second or longer.  Though we now have
  proof of functional heterogeneity in a handful of systems---enzymes,
  motors, adhesion complexes---identifying and measuring it remains a
  formidable challenge.  Here we show that evidence of this phenomenon
  is more widespread than previously known, encoded in data collected
  from some of the most well-established single molecule techniques:
  AFM or optical tweezer pulling experiments.  We present a
  theoretical procedure for analyzing distributions of
  rupture/unfolding forces recorded at different pulling speeds.  This
  results in a single parameter, quantifying the degree of
  heterogeneity, and also leads to bounds on the equilibration and
  conformational interconversion timescales.  Surveying 
    ten published datasets, we find heterogeneity in five of them,
  all with interconversion rates slower than 10 s$^{-1}$.  Moreover,
  we identify two systems where additional data at realizable pulling
  velocities is likely to find a theoretically predicted, but so far
  unobserved cross-over regime between heterogeneous and
  non-heterogeneous behavior.  The significance of this regime is that
  it will allow far more precise estimates of the slow conformational
  switching times, one of the least understood aspects of functional
  heterogeneity.
\end{abstract}

\maketitle

\section*{Introduction}

One of the great problems in modern biology is to understand how the
intrinsic diversity of cellular behaviors is shaped by factors outside
of the genome.  The causes of this heterogeneity are spread across
multiple scales, from noise in biochemical reaction networks through
epigenetic mechanisms like DNA methylation and histone
modification~\cite{Altschuler2010}.  It might be natural to expect
heterogeneity at the cellular level because of the bewildering array
of time and length scales associated with the molecules of life that
govern cell function. Surprisingly, even at the level of individual
biomolecules, diversity in functional properties like rates of
enzymatic catalysis~\cite{Lu1998,Oijen2003,English2005,Solomatin2010}
or receptor-ligand binding~\cite{Kim2010, Buckley2014} can occur.
 This diversity arises from the presence of many distinct
  functional states in the free energy landscape, which correspond to
  long-lived active conformations of the biomolecule.  Though the
reigning paradigm in proteins and nucleic acids has been a single,
folded native structure, well separated in free energy from any other
conformations, possibilities about rugged landscapes with multiple
native states have been explored for a long
time~\cite{Austin1975,Frieden1979,Schmid1981,Agmon1983,Frauenfelder1988,Honeycutt1990,Zwanzig1990,Zwanzig1992}.
Yet only with the revolutionary advances in single molecule
experimental techniques in recent years have we been able to gather
direct evidence of functional heterogeneity, in systems ranging from
protein enzymes~\cite{Lu1998,Oijen2003,English2005} and nucleic
acids~\cite{Solomatin2010,Hyeon2012,Kowerko15PNAS}, to molecular
motors~\cite{Liu2013} and cell adhesion complexes~\cite{Kim2010,
  Buckley2014}.  As research inevitably moves toward larger
macromolecular systems, the examples of functional heterogeneity will
only multiply.  We thus need to develop theories that can deduce
aspects of the hidden kinetic network of states underlying the single
molecule experimental data~\cite{Presse13}, allowing us to quantify
the nature and extent of the heterogeneity.

The focus in this study is single molecule force spectroscopy,
conducted either by AFM or optical tweezers, which constitutes an
extensive experimental literature over the last two decades.  Our
contention is that evidence of heterogeneity is widespread in this
literature, but has gone largely unnoticed, since researchers (with a
few exceptions, discussed
below~\cite{Raible2004,Raible2006EPL,Raible2006BJ,Hyeon2014}) did not
recognize the markers in their data that indicated heterogeneous
behavior.  To remedy this situation, we introduce a universal approach
to analyzing distributions of rupture/unfolding forces collected in
pulling experiments, which yields a single non-dimensional parameter
$\Delta \ge 0$.  The magnitude of $\Delta$ characterizes the extent of
the disorder in the underlying ensemble, the ruggedness of the free
energy landscape.  Moreover, our method provides a way of estimating
bounds on key timescales, describing both the fast local equilibration
in each well (distinct system state) of our rugged landscape, and the
slow interconversion between the various wells.  After verifying the
validity of our approach using synthetic data generated from a
heterogeneous model system, we survey ten experimental datasets,
comprising a diverse set of biomolecular systems from simple DNA
oligomers to large complexes of proteins and nucleic acids.  The
largest values of $\Delta$ in our survey, indicating the strongest
heterogeneity, come from systems involving nucleic acids alone or
protein/nucleic acid interactions, supporting the hypothesis that
nucleic acid free energy landscapes are generally more rugged than
those involving only proteins~\cite{Thirum05Biochem}.  Our theory thus
provides a powerful new analytical tool, for the first time allowing a
broad comparison of functional heterogeneity among different
biomolecules through a common experimental protocol.

\section*{Theory}

\begin{figure}
\centering\includegraphics[width=0.95\columnwidth]{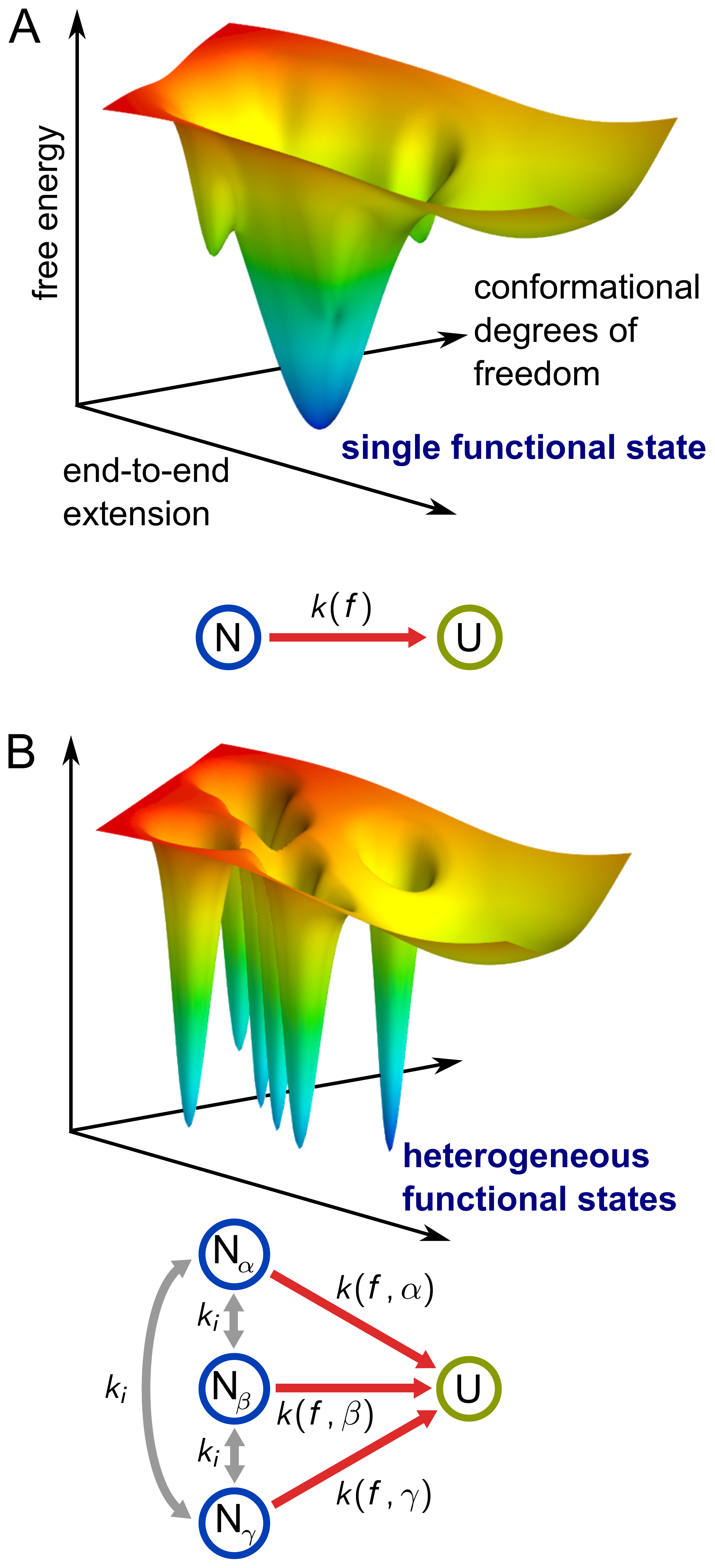}
\caption{A) Schematic biomolecular free energy landscape with a single
  functional state, N, corresponding to an ensemble of folded/bound
  conformations.  Under an adiabatically increasing external force
  {$f(t)$}, there is a instantaneous rupture rate {$k(f(t))$} describing
  transitions between N and the unfolded/unbound ensemble U.  B)
  Schematic free energy landscape of a heterogeneous system with
  multiple functional states.  Each functional ensemble $N_\alpha$ will
  have a state-dependent adiabatic rupture rate {$k(f,\alpha)$}.
  Assuming the states are roughly equally probable in equilibrium,
  there will be a single overall rate $k_\text{i}$ for interconversion
  between the various states.}\label{land}
\end{figure}

{\em Force spectroscopy for a pure, adiabatic system:} As a starting
point, consider a generic free energy landscape for a biomolecular
system with a single functional state (Fig.~1A) subject to an
increasing time-dependent external force $f(t)$.  For a molecular
complex, the functional basin of attraction in the landscape would
correspond to an ensemble of bound conformations with similar
energies, which we label N.  For the case of single molecule folding,
this would be the unique native ensemble.  The force is applied
through an experimental apparatus like an AFM or optical tweezer,
typically connected to the biomolecule through protein or nucleic acid
linkers of known stiffness.  The apparatus is pulled at a constant
velocity $v$, leading to a force ramp with slope $df/dt = \omega_s(f)v$,
where $\omega_s(f)$ is the effective stiffness of the setup (linkers plus
the AFM cantilever or optical trap).  This $\omega_s(f)$ may in general
depend on the force, particularly for the AFM setup, where the
cantilever stiffness is often comparable to or greater than that of
the molecular construct.  So we also define a characteristic stiffness
$\bar{k}_s$ which we set to the mean $\omega_s(f)$ over the range of forces
probed in the experiment (though the precise value of $\bar{k}_s$ is
not important).  This allows us to introduce a characteristic force
loading rate $r$ proportional to the velocity, $r = \bar{k}_sv$.

If at time $t=0$ the system starts in N, the force ramp tilts the
landscape along the extension coordinate.  If we model the
conformational dynamics of the system as diffusion within this
landscape, the tilting eventually leads to a transition out of N,
associated with unbinding of the complex or unfolding of the molecule
(an ensemble of states we call U).  We let $\Sigma_r(t)$ be the
survival probability for loading rate $r$, in other words the
probability that the transition to U has not occurred by time $t$.
The distribution of first rupture times is then $-d\Sigma_r/dt$, and
the mean rupture rate $\bar{k}(r)$ is just the inverse of the average
rupture time,
\begin{equation}\label{e1}
\bar{k}(r) =\left[\int_0^\infty dt\,t \left(-\frac{d\Sigma_r}{dt}\right)\right]^{-1} = \left[\int_0^\infty dt\,\Sigma_r(t)\right]^{-1},
\end{equation}
where we have used integration by parts and assumed that rupture always
occurs if we wait long enough, $\Sigma(\infty) = 0$.

The behavior of $\Sigma_r(t)$ at different $r$ depends on how
$\bar{k}(r)$ compares to two other intrinsic rates.  The first is the
equilibration rate $k_\text{eq}$ in the N well, or how quickly the
system samples the configurations of the functional ensemble.  For a
single, smooth well with mean curvature $\omega_0$ and a diffusion
constant $D$, this rate is on the order of
$k_\text{eq} \sim \beta \omega_0 D$, where $\beta = 1/k_B T$.  The
second is a critical rate $k_\text{c}(r) = r/f_c$, which describes how
quickly the force reaches a critical force scale for rupture
$f_c \sim G^\ddagger/x^{\ddagger}$.  Here $G^\ddagger$ is the energy
scale of the barrier that needs to be overcome for the N to U
transition at zero force, and $x^{\ddagger}$ the extension difference
between the N well minimum and the transition state.  For
$f \gtrsim f_c$ the landscape is tilted sufficiently that the barrier
becomes insignificant, and rupture occurs quickly (on a
diffusion-limited time scale).  If
$k_\text{c}(r) \ll \bar{k}(r) \ll k_\text{eq}$, the system is in the
adiabatic regime.  The force ramp is sufficiently slow that rupture
occurs before the critical force is reached, and equilibration is fast
enough that the system can reach quasi-equilibrium at the
instantaneous value of the force $f(t)$ at all times $t$ before the
rupture.

If the adiabatic condition is satisfied, the survival probability
$\Sigma_r(t)$ obeys the kinetic equation
$d\Sigma_r(t)/dt = -k(f(t)) \Sigma_r(t)$, where $k(f)$ is the rupture
rate at constant force $f$.  Since {$f(t)$} is a monotonically
increasing function of $t$, we can change variables from $t$ to
{$f(t)$}~\cite{Evans1997}, and solve for {$\Sigma_r(f)$}, the
probability that the system does not rupture before the force value
{$f$} is reached: {\begin{equation}\label{e2} \Sigma_r(f) =
    \exp\left(-\frac{1}{r} \int_0^f df^\prime \frac{\bar{k}_s
        k(f^\prime)}{\omega_s(f^\prime)}\right).
\end{equation}
}
Interestingly, the integral inside the exponential is independent of
the loading rate $r$.  Hence for a system pulled from a single native
ensemble, we can calculate the following quantity from experimental trajectories at different $r$,
{
\begin{equation}\label{e3}
\Omega_r(f) \equiv -r \log \Sigma_r(f),
\end{equation}
} and the results should collapse onto a single master curve for all
$r$ in the adiabatic regime.  When $r$ is sufficiently large that
$\bar{k}(r) < k_\text{c}(r)$ or {$\bar{k}(r) > k_\text{eq}$}, the
assumption of quasi-equilibrium on a slowly changing energy landscape
breaks down, and Eq.~\eqref{e2} no longer holds.  For this fast,
non-adiabatic case~\cite{Hu2010,Bullerjahn2014} we should find that
{$\Omega_r(f)$} varies with $r$, as we will explore later in more
detail.

\vspace{1em}

{\em Force spectroscopy for a heterogeneous, adiabatic system:} In a
pioneering series of studies, Raible and collaborators analyzed force
ramp experiments for the regulatory protein ExpG unbinding from a DNA
fragment~\cite{Raible2004,Raible2006EPL,Raible2006BJ}.  Plotting
{$\Omega_r(f)$} (the data reproduced in Fig.~6D), they did not
find any collapse, as might be surmised from Eq.~\ref{e3}.  
This was not an artifact due to non-adiabaticity
(violation of the inequality
{$k_\text{c}(r) \ll \bar{k}(r) \ll k_\text{eq}$}), since the absence of
collapse becomes even more pronounced at small loading rates, further
into the adiabatic territory where collapse should be observed.  They
correctly surmised that the cause of this divergence is heterogeneity
in the ensemble of states in the protein-DNA complex.

To understand the behavior of {$\Omega_r(f)$} in a heterogeneous
system, let us consider the effects of a force ramp on a biomolecular
free energy landscape with multiple functional states
(Fig.~1B).  Our goal is to use {$\Omega_r(f)$}, derived from
experimental pulling trajectories, to quantify the extent of the
heterogeneity and extract information about the underlying
conformational dynamics.  The functional states are distinct basins of
attraction in the landscape, corresponding to distinct functional
ensembles which we label N$_\alpha$ for state $\alpha$.  We assume the
minimum energy in each well and their overall dimensions are
comparable, so that the equilibrium probabilities $p^\text{eq}_\alpha$
of the various states are of the same order.  In this case if
$\alpha \ne \alpha^\prime$, the transition rates
$k_{\alpha\to \alpha^\prime}$ and $k_{\alpha^\prime \to\alpha}$ are
also similar from detailed balance,
$k_{\alpha \to \alpha^\prime} / k_{\alpha^\prime \to \alpha} =
p^\text{eq}_{\alpha^\prime}/p^\text{eq}_\alpha \sim \text{O}(1)$.
Hence we can introduce an overall scale for the interconversion rate
between the different states, $k_\text{i}$, such that
$k_{\alpha \to \alpha^\prime} \sim \text{O}(k_\text{i})$ for any
$\alpha \ne \alpha^\prime$.  Thus we now have two intrinsic time
scales: {$k_\text{eq}$} for equilibration within a single N$_\alpha$,
and $k_\text{i}$ for transitions between distinct N$_\alpha$'s, where
typically {$k_\text{i} < k_\text{eq}$}.

The experimental setup is the same as above, with a loading rate $r$,
and a corresponding mean rupture rate $\bar{k}(r)$ for reaching the U
ensemble.  We can identify three dynamical regimes, based on the
magnitude of $k_\text{i}$.  In the first regime, interconversion is
slow, with $k_\text{i} \ll \bar{k}(r)$.  In the second regime,
$k_\text{i}$ is comparable to $\bar{k}(r)$.  In fact, as we will
discuss later in more detail, we will be particularly interested in
the cross-over scenario where $k_\text{i}\ge \bar{k}(r)$ for some
subset of the $r$ values in the experiment, but
$k_\text{i} < \bar{k}(r)$ for the remainder.  If this second regime is
identified in an experiment, it provides a way to estimate the scale
of $k_\text{i}$.  Finally, in the third regime, the barriers between
the N$_\alpha$ basins of attraction are small, such that
$k_\text{i} \gg \bar{k}(r)$, and the system can sample all the states
before rupture.  Qualitatively, this scenario is indistinguishable
from the case of a system with a single native basin of attraction,
with $k_\text{i}$ taking the role of {$k_\text{eq}$} as the rate scale
for overall equilibration in the landscape.  Since the first regime is
simpler to treat mathematically than the second regime, we will
initially focus on a theory to describe the first regime and identify
its signatures in experimental data.  Assessing the validity of this
theory in experiments will turn out to be a useful criterion for
distinguishing between the first, second, and third regimes, and thus
putting bounds on $k_\text{i}$. This byproduct of our theory is of
considerable importance because it is {\it a priori} very difficult to
estimate $k_\text{i}$.

To begin, consider adiabatic pulling where $k_\text{i}$ is the slowest
rate in the system,
{$k_\text{i} \ll k_\text{c}(r) \ll \bar{k}(r) \ll k_\text{eq}$}.  On
the time scale of pulling and rupture, the system is effectively
trapped in a heterogeneous array of states: if we start a pulling
trajectory in state $\alpha$, the system will remain in that state
until rupture.  The rupture rate at constant force, {$k(f,\alpha)$}
will in general depend on the state, and the ensemble of molecules
from which we pull will be characterized by a set of initial state
probabilities $p_\alpha$.  If $k_\text{i}$ is extremely small, such
that the system cannot interconvert even on the macroscopic time
scales of experimental preparation, $p_\alpha$ may be different from
$p^\text{eq}_\alpha$, since we are not guaranteed to draw from an
equilibrium distribution across the entire landscape.  This
distinction is not important for the analysis below.  In fact, our
approach also works when $k_\text{i} = 0$, corresponding to the
quenched disorder limit, as seen for example in an ensemble of
molecules with covalent chemical differences.

The analogue of Eq.~\ref{e2} for the survival probability
{$\Sigma_r(f)$} during adiabatic pulling in a heterogeneous system
with small $k_\text{i}$ is
{\begin{equation}\label{e4}
\Sigma_r(f) = \left\langle\exp\left(-\frac{1}{r} \int_0^f df^\prime \frac{\bar{k}_s k(f^\prime,\alpha)}{\omega_s(f^\prime)}\right)\right\rangle,
\end{equation}
}
where the brackets denote an average over the initial ensemble of
states, $\langle O(\alpha) \rangle \equiv \sum_\alpha p_\alpha
O(\alpha)$ for any quantity $O(\alpha)$.  
The associated {$\Omega_r(f)$}
from Eq.~\eqref{e3} can be expressed through a cumulant expansion in
terms of the integrand {$I(f,\alpha) \equiv \int_0^f df^\prime
\bar{k}_s k(f^\prime,\alpha)/\omega_s(f^\prime)$} as follows:
{
\begin{equation}\label{e5}
\begin{split}
\Omega_r(f) &= -\sum_{n=1}^\infty (-1)^n \frac{\kappa_n(f)}{n! r^{n-1}},\\
\kappa_n(f) &\equiv \left.\frac{\partial^n}{\partial \lambda^n} \log \langle e^{\lambda I(f,\alpha)} \rangle\right|_{\lambda=0}.
\end{split}
\end{equation}
}
The first two cumulants are 
{$\kappa_1(f) = \langle I(f,\alpha)\rangle$}
and 
{$\kappa_2(f) = \langle I^2(f,\alpha) \rangle - \langle I(f,\alpha)
\rangle^2$}.  
In the absence of heterogeneity, all cumulants
{$\kappa_n(f)$} with $n>1$ are exactly zero.  
For a small degree of
heterogeneity, or equivalently for sufficiently fast loading rates
$r$, the main contribution to the expansion is from the $n=1$ and
$n=2$ terms.  For the case of fast $r$ we assume that we are still
within the adiabatic regime, where $k_\text{c}(r) \ll \bar{k}(r)$, which
turns out to be valid even for the largest loading rates in the
experimental studies discussed below.  In this scenario, where the
$n>2$ contributions are negligible, {$\Omega_r(f)$} can be approximated
as
{\begin{equation}\label{e6}
\Omega_r(f) \approx \frac{r}{\Delta(f)} \log \left(1+\frac{\kappa_1(f)\Delta(f)}{r}\right)
\end{equation}
}
where {$\Delta(f) \equiv \kappa_2(f)/\kappa^2_1(f) \ge 0$} is a
dimensionless measure of the ensemble heterogeneity.  For a pure
system, {$\Delta(f) \to 0$}, giving {$\Omega_r(f) \to \kappa_1(f)$},
independent of $r$.  Eq.~\eqref{e6} agrees with the expansion in
Eq.~\eqref{e5} up to order $n=2$, and also has the nice property that
it satisfies the inequality {$\Omega_r(f) \le \kappa_1(f)$}, just like
the exact form.  The latter inequality follows from the definition of
{$\Sigma_r(f)$} in Eq.~\eqref{e4} and Jensen's inequality, 
{$\Sigma_r(f)
\ge \exp(-\kappa_1(f)/r)$}.

{\em Implementing the model on experimental data:} So far the
discussion has been completely general, but to fit Eq.~\eqref{e6} to
experimental data we need specific forms for {$\Delta(f)$} and
{$\kappa_1(f)$}.  The minimal physically sensible approximation, with
the smallest number of unknown parameters, supplements Eq.~\eqref{e6}
with the assumptions,
\begin{equation}\label{e7}
\Delta(f) = \Delta, \qquad \kappa_1(f) = \frac{k_0}{\beta x^{\ddagger}} \left(e^{\beta f x^{\ddagger}}-1\right).
\end{equation}
The constants $\Delta$, $k_0$, and {$x^{\ddagger}$} are fitting
parameters.  This presumes that {$\Delta(f)$} changes little over the
range of forces in the data, and {$\kappa_1(f)$} has the same
mathematical form as in a pure Bell model with an escape rate
{$k(f) = k_0 e^{\beta f x^{\ddagger}}$} and
{$\omega_s(f) = \bar{\omega}_s$}, where $k_0$ is the escape rate at
zero force and {$x^{\ddagger}$} the distance to the transition state.
For a heterogeneous system, the parameters $k_0$ and {$x^{\ddagger}$}
no longer have this simple interpretation, but we can still treat them
as effective Bell values, averaged over the ensemble, with $\Delta$
measuring the overall scale of the heterogeneity.  Eq.~\eqref{e6},
together with the three-parameter approximation of Eq.~\eqref{e7},
provides remarkably accurate fits to all the heterogeneous
experimental data sets we have encountered in the literature.  As will
be seen below, it is capable of simultaneously fitting {$\Omega_r(f)$}
data for loading rates $r$ spanning nearly two orders of magnitude.

Though we focus on $\Omega_r(f)$ as the main
  experimental quantity of interest, Eqs.~\eqref{e6}-\eqref{e7} can
  also be used to derive a closed form expression for the probability
  distribution of rupture forces,
  $p_r(f) = -d\Sigma_r(f)/df = -(d/df) \exp(-\Omega_r(f)/r)$, at
  loading rate $r$:
\begin{equation}\label{pr}
p_r(f) =\frac{k_0 e^{\beta f x^\ddagger}}{r}\left(1+\frac{\Delta k_0 (e^{\beta f x^\ddagger}-1)}{\beta r x^\ddagger} \right)^{-\frac{\Delta+1}{\Delta}}.
\end{equation}
In the limit of no heterogeneity, $\Delta \to 0$, this distribution
reduces to the one predicted for a Bell model under a constant loading
rate~\cite{Evans1997}.  The theoretical form for $p_r(f)$ also allows
us to carry out a relative likelihood analysis on the experimental
data, to verify that $\Delta$ is indeed a robust indicator of
heterogeneity.  As detailed in SI Sec.~6, we found that experimental
distributions $p_r(f)$ corresponding to systems with nonzero $\Delta$
were far more likely to be described by the heterogeneous theory in
Eq.~\eqref{pr} than a pure model with the same number of parameters.
We surmise that if analysis of experimental data using our theory
indicates that $\Delta \ne 0$ then it is highly probable that any
single state model is insufficient to describe the system, and a
multiple state description is needed.

To verify that our analysis and conclusions would not change
substantially if the assumptions of the minimal model were relaxed, we
have also tested two generalized versions of the model: one using the
Dudko-Hummer-Szabo~\cite{Dudko2006} instead of the Bell form for the
escape rate in $\kappa_1(f)$, and the other allowing $\Delta(f)$ to
vary linearly with $f$ across the force range.  Both extensions have
four instead of three fitting parameters, but the heterogeneity
results for the experimental systems we analyzed are completely
consistent with those obtained using the minimal model (see
Supplementary Information (SI) Sec.~1 for details).  These results
demonstrate that if the need arises in future experimental contexts,
the theory leading to Eq.~\eqref{e6} is quite general, and can be
tailored by choosing suitable expressions for $\kappa_1(f)$ and
$\Delta(f)$ that go beyond the minimal model of Eq.~\eqref{e7}.

The theory described up to now applies only to the first dynamical
regime, where $k_\text{i} \ll \bar{k}(r)$.  However the cases where
$k_\text{i}$ is larger than some or all of the $\bar{k}(r)$, and the
theory partially or completely fails, turn out to be very informative
as well.  To understand these points, it is easier to discuss the
theory in the context of a concrete physical model for heterogeneity,
which we introduce in the next section.

\vspace{1em}

\section*{Results and Discussion}

\begin{figure*}
\centering\includegraphics[width=0.9\textwidth]{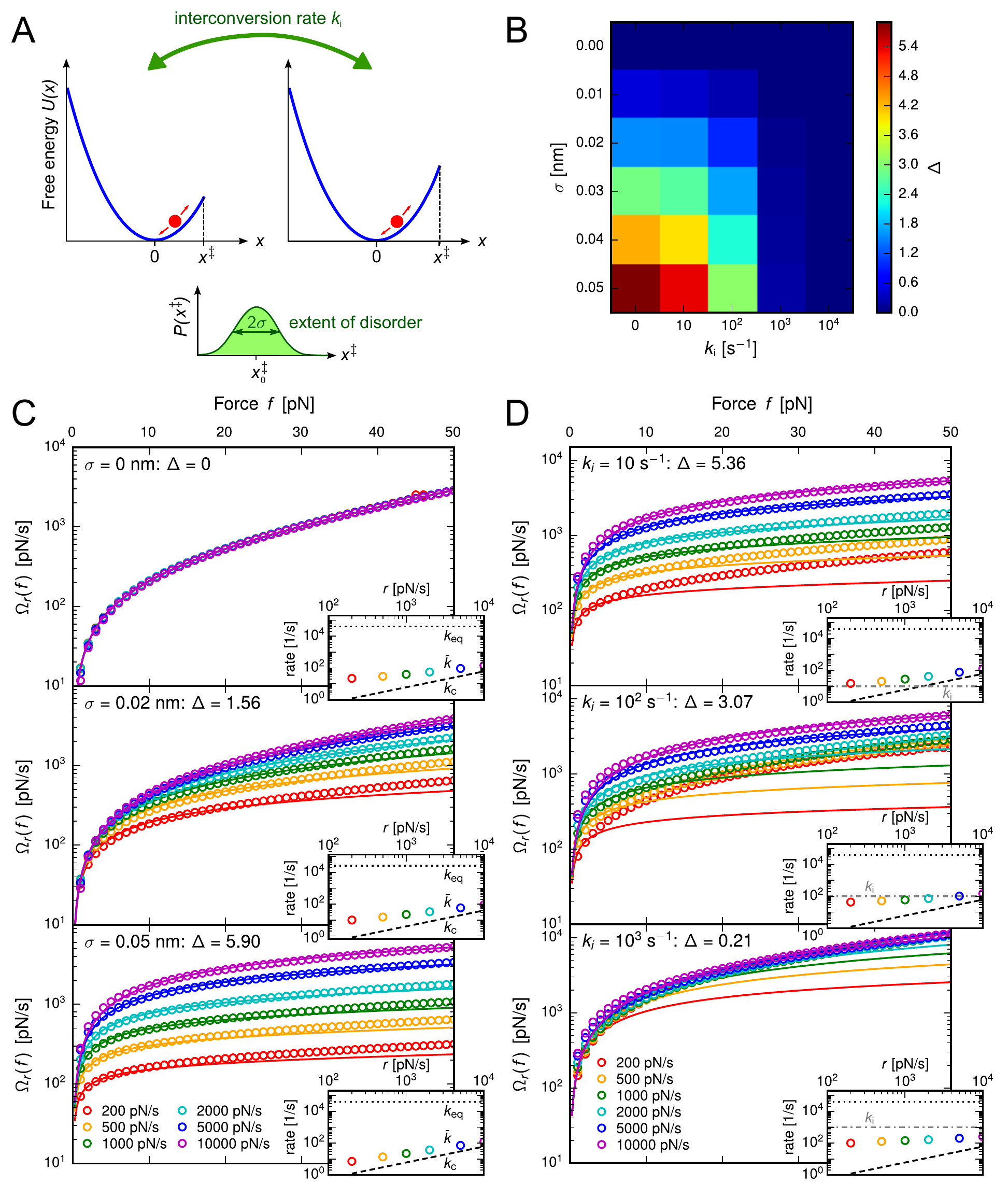}
\caption{Analysis of the FBL heterogeneous model system.  A) Two
  different free energy wells, corresponding to distinct states,
  characterized by different transition distances to rupture,
  {$x^{\ddagger}$} and {$(x^{\ddagger})^\prime$}.  The system switches
  to a new value of {$x^{\ddagger}$}, drawn from a Gaussian
  distribution centered at {$x^{\ddagger}_0$} with standard deviation
  $\sigma$, with rate $k_\text{i}$.  B) Heat map of $\Delta$ as it
  varies with $\sigma$ and $k_\text{i}$, extracted from fitting the
  theory of Eqs.~\eqref{e6}-\eqref{e7} to numerical simulation results
  of {$\Omega_r(f)$} for the model system.  The parameters are:
  $D = 100$ nm$^2$/s, $\omega_0 = 400$ $k_BT/$nm$^2$,
  {$x^{\ddagger}_0=0.2$} nm, $\sigma = 0-0.05$ nm,
  $k_\text{i} = 0 - 10^4$ s$^{-1}$.  C-D) Sample simulation results
  {$\Omega_r(f)$} (circles) on a logarithmic scale, with each color
  denoting a different loading rate $r$.  The panels show different
  combinations of $k_\text{i}$ and $\sigma$, with the plots in C
  illustrating the case of quenched disorder ($k_\text{i} = 0$) for
  increasing $\sigma$, and D showing increasing $k_\text{i}$ for fixed
  $\sigma = 0.05$.  The theoretical best-fit curves are drawn as solid
  curves, and the resulting $\Delta$ value is listed in each plot.
  The insets show the mean rupture rate $\bar{k}(r)$ (circles) as a
  function of $r$ compared to $k_\text{eq}$ (dotted line),
  $k_\text{c}(r)$ (dashed line), and $k_\text{i}$ (dash-dotted
  line).}\label{toy}
\end{figure*}

{\em Fluctuating Barrier Location (FBL) model:} Before turning to
experimental data, we verify that the $\Delta$ parameter extracted
from the fitting of {$\Omega_r(f)$} curves using
Eqs.~\eqref{e6}-\eqref{e7} is a meaningful measure of heterogeneity.
To do this, we will generate synthetic rupture data from a
heterogeneous model system.  The FBL model, illustrated in
Fig.~2A, consists of a reaction coordinate $x$ whose dynamics
are described by diffusion with constant $D$ along a parabolic free
energy $U(x) = (1/2) \omega_0 x^2$ for {$x\le x^{\ddagger}$}.  Rupture
occurs if $x$ exceeds the transition distance {$x^{\ddagger}$}.  To
mimic dynamic heterogeneity, the value of {$x^{\ddagger}$} changes at
random intervals, governed by a Poisson process with an
interconversion rate $k_\text{i}$.  At every switching event, a new
value of {$x^{\ddagger}$} is drawn from a Gaussian probability
distribution
{$P(x^{\ddagger}) =
  \exp(-(x^{\ddagger}-x^{\ddagger}_0)^2/2\sigma^2)/\sqrt{2\pi
    \sigma^2}$} centered at {$x^{\ddagger}_0$} with standard deviation
$\sigma$, and diffusion continues if $x$ is less than the transition
distance.  At time $t=0$, when the applied force ramp {$f(t) = rt$}
begins, we assume the initial ensemble of systems all start at $x=0$
with {$x^{\ddagger}$} values distributed according to
{$P(x^{\ddagger})$}.  Survival probabilities {$\Sigma_r(f)$} are
computed from numerical simulations of the diffusive process, with
about $3\times 10^4$ rupture events collected for each parameter set
(see the SI Sec.~2 for additional
details).  The simplicity of the model, where one parameter, $\sigma$,
controls the degree of heterogeneity, and another, $k_\text{i}$, the
interconversion dynamics, allows us to explore the behavior of
{$\Sigma_r(f)$}, and hence {$\Omega_r(f)$}, over a broad range of
disorder and intrinsic time scales.

The circles in Fig.~2C-D show simulation results for
{$\Omega_r(f)$} between {$f= 0 - 50$} pN, plotted on a logarithmic scale,
with each color denoting a different ramp rate in the range
$r=200-10000$ pN/s.  The model parameters are $D = 100$ nm$^2$/s,
$\omega_0 = 400$ $k_BT/$nm$^2$, {$x^{\ddagger}_0=0.2$} nm, $\sigma = 0-0.05$ nm,
$k_\text{i} = 0 - 10^4$ s$^{-1}$, which give a variety of
{$\Omega_r(f)$} curves of comparable magnitude over similar force scales
to the experimental data discussed below.  Fig.~2C shows
results for quenched disorder ($k_\text{i} = 0$) at different
$\sigma$, while Fig.~2D shows results for varying $k_\text{i}$
at fixed $\sigma = 0.05$ nm.  For a given choice $k_\text{i}$ and
$\sigma$, we fit the analytical form of Eq.~\eqref{e6}-\eqref{e7}
simultaneously to the six {$\Omega_r(f)$} curves at different $r$, with
the best-fit model plotted as solid lines in Fig.~2C-D.  This
fitting yields values for $\Delta$, $k_0$, and {$x^{\ddagger}$} in each case.  The
variation of $\Delta$ with $\sigma$ and $k_\text{i}$ is plotted as a
heat map in Fig.~2B.

Let us first consider the quenched disorder results (Fig.~2C and
the left column of Fig.~2B).  By definition, since $k_\text{i}
= 0$, the system ensemble is permanently frozen in a heterogeneous
array of different states with different values of {$x^{\ddagger}$}.  Moreover, the
adiabatic assumptions also hold, as can be seen in the insets to
Fig.~2C.  These show the mean rupture rate $\bar{k}(r)$ for
different $r$ (circles) compared to {$k_\text{eq}$} (dotted line) and
$k_\text{c}(r)$ (dashed line).  For all the $r$ values analyzed,
{$k_\text{c}(r) < \bar{k}(r) \ll k_\text{eq}$}, so adiabaticity should
approximately hold.  Thus the assumptions leading to
Eqs.~\eqref{e6}-\eqref{e7} are valid, and indeed the analytical form
provides an excellent fit to the simulation data.  Though the theory is by
construction most accurate in the limit of fast (but still adiabatic)
$r$, it still quantitatively describes the results for $r$ spanning
two orders of magnitude.  Only small discrepancies start to appear at
the slowest loading rates.  For the pure system limit ($\sigma=0$) the
best-fit value of $\Delta$ is also zero, with all the {$\Omega_r(f)$}
curves collapsing on one another.  $\Delta$ progressively increases
with $\sigma$, growing roughly proportional to the width of the
disorder distribution.  The greater the heterogeneity, the more
pronounced the separation between the {$\Omega_r(f)$} curves at various
$r$.

The results in Fig.~2D are obtained by keeping the extent of
heterogeneity fixed at a large level ($\sigma = 0.05$ nm) and allows
interconversion, increasing $k_\text{i}$ from 10 to $10^3$ s$^{-1}$.
So long as $\bar{k}(r) \gg k_\text{i}$, the system is unlikely to
interconvert on the time scale of rupture, and we see distinct,
non-collapsed {$\Omega_r(f)$} curves.  But as $k_\text{i}$ increases
and overtakes $\bar{k}(r)$, starting from the smallest values of $r$
where $\bar{k}(r)$ has the smallest magnitude, the {$\Omega_r(f)$}
curves begin to collapse on one another. This leads to increasing
discrepancies between the data and the theoretical fit, since the
assumptions justifying the theory break down when
$\bar{k}(r) < k_\text{i}$.  Eventually, once $k_\text{i}$ is greater
than all the $\bar{k}(r)$, there is total collapse of the
{$\Omega_r(f)$} curves (bottom panel of Fig.~2D).  Frequent
interconversion between the different states of the system before
rupture averages out the heterogeneity, making the results
indistinguishable from a pure system.   In this limit the
  ensemble of functional states acts effectively like a single
  functional basin of attraction, with multiple distinct pathways to
  rupture.  Though multiple pathways between a pair of states can be
  considered to be another manifestation of heterogeneity, they are not in themselves
  sufficient to lead to non-collapse of the $\Omega_r(f)$ curves, as
  we discuss in more detail in SI Sec.~3.  To see anything but
  complete collapse of the $\Omega_r(f)$ curves in the adiabatic
  regime requires a small enough interconversion rate $k_\text{i}$,
  slower than the mean rupture rates $\bar{k}(r)$ for at least some
  subset of the $r$ values.

\vspace{1em}

  {\em Dynamical regimes and extraction of  bounds on 
    time scales of internal dynamics:} Interestingly, it is precisely the discrepancy in
  the theoretical fits with increasing $k_\text{i}$ that points the
  way to one of the most valuable features of our approach.  Not only
  can we measure heterogeneity, but also infer information about the
  time scales of conformational dynamics.  Note first that the
  best-fit values of $\Delta$ track the disappearance of
  heterogeneity, monotonically decreasing from $\Delta = 5.90$ at
  $\sigma=0.05$, $k_\text{i} = 0$ s$^{-1}$, to $\Delta = 0.21$
  at $\sigma=0.05$, $k_\text{i} = 10^3$ s$^{-1}$.  It clear
  however that as $k_\text{i}$ increases and dynamical disorder
  becomes more prominent, a single overall value of $\Delta$ is an
  imperfect description of the dynamics.  We can get a more
  fine-grained picture by looking at $\Delta$ calculated from smaller
  subsets of the data, and how it varies with the mean time scale of
  rupture $\bar{k}$.  To do this let us take $\Omega_r(f)$ curves from
  two consecutive loading rates $(r_1, r_2)$, fit
  Eq.~\eqref{e6}-\eqref{e7}, and calculate the resulting value of
  $\Delta$, which we will call the ``pair'' parameter
  $\Delta_\text{p}(r_1,r_2)$.  For example, if our total data set consists of
  six loading rates $r = 200$, 500, 1000, 2000, 5000, 10000 pN/s, we
  first do this for $(r_1,r_2) = (200,\:500)$ pN/s, then
  $(500,\:1000)$ pN/s, and so on, to get five different results for
  $\Delta_\text{p}(r_1,r_2)$.  The advantage of this approach is that each
  $\Delta_\text{p}$ corresponds to a much smaller range of rupture time
  scales than what is covered by the entire data set.  In
  Fig.~3A we plot $\Delta_\text{p}$ for $\sigma=0.05$,
  $k_\text{i} = 0$, 10, 10$^2$, 10$^3$ s$^{-1}$.  The $x$-axis
  coordinate is the smaller mean rupture rate of the pair,
  $\bar{k} = \text{min}(\bar{k}(r_1),\bar{k}(r_2))$.

  The behavior of $\Delta_\text{p}$ in Fig.~3A allows us to identify
  three different behaviors, corresponding to the three
    dynamical regimes discussed in the Theory section:

\begin{enumerate}

 \item {\it Non-collapse (NC):} Here all the
  $\Delta_\text{p}(r_1,r_2) \ge 1$, and $\Delta_\text{p}(r_1,r_2)$ for
  any pair of $(r_1,r_2)$ is approximately the same as $\Delta$
  calculated from the entire data set.  We see this in the
  $k_\text{i} = 0$ s$^{-1}$ case in Fig.~3A, where for comparison the
  value of $\Delta$ over the whole set is marked by a horizontal
  dashed line.  The corresponding $\Omega_r(f)$ curves are in the
  bottom panel of Fig.~2C.  The agreement between
  $\Delta_\text{p}(r_1,r_2)$ and $\Delta$ is a consistency check for
  the theory, and implies that the underlying assumptions are valid,
  namely $k_\text{i} < \bar{k}(r)$ and $k_\text{eq} > \bar{k}(r)$ for
  all $r$ in the data set.  From this we can conclude that the minimum
  value of $\bar{k}(r)$ among all the loading rates $r$ used in the
  experiment gives us an upper bound on $k_\text{i}$.  Similarly the
  maximum value of $\bar{k}(r)$ over all $r$ gives a lower bound on
  $k_\text{eq}$.  For $k_\text{i} = 10$ s$^{-1}$ in Fig.~3A, we see
  what happens as $k_\text{i}$ approaches the time scale of
  $\bar{k}(r)$.  We are still in the NC regime, since
  $\Delta_\text{p} \ge 1$ and $k_\text{i}$ (vertical dotted line) is
  smaller than any of the $\bar{k}(r)$.  But $k_\text{i}$ is now
  sufficiently close to $\bar{k}(r=200\:\text{pN/s})$ that
  $\Delta_\text{p}(200,500)$ (the leftmost point) is smaller than the
  rest of the $\Delta_\text{p}$, which lie at faster rupture
  timescales relatively unaffected by $k_\text{i}$.

\item {\it Partial collapse (PC):} $\Delta_\text{p}(r_1,r_2) \ge 1$ for
  the largest values of $(r_1,r_2)$, but for small loading rates
  $\Delta_\text{p}(r_1,r_2) \ll 1$.  This occurs in the
  $k_\text{i} = 10^2$ s$^{-1}$ results in Fig.~3A.  In this
  regime the system is adiabatic, $k_\text{eq} > \bar{k}(r)$, but now
  $k_\text{i}$ falls between the smallest and largest values of
  $\bar{k}(r)$.  In the $k_\text{i} = 10^2$ s$^{-1}$ case, the
  variation in $\Delta_\text{p}$ is a reflection of the degree of
  overlap in the $\Omega_r(f)$ curves (middle panel of
  Fig.~2D).  The $(r_1,r_2) = (5000,10000)$ pN/s pair (blue
  and purple $\Omega_r(f)$ circles) are clearly separated,
  corresponding to $\Delta_\text{p} \ge 1$ and the fact that
  $k_\text{i} \lesssim \bar{k}(r_1)$, $\bar{k}(r_2)$.  The $(200,500)$
  pN/s pair (red and orange circles) are nearly overlapping,
  corresponding to $\Delta_\text{p} \ll 1$, and
  $k_\text{i} > \bar{k}(r_1)$, $\bar{k}(r_2)$.  The PC regime thus
  provides the best case scenario for directly estimating $k_\text{i}$
  from the data, since we can bound $k_\text{i}$ from above and below,
  and we know $k_\text{i}$ will roughly coincide with the $\bar{k}$
  where $\Delta_\text{p}(r_1,r_2) \sim 1$.

\item {\it Total collapse (TC):} $\Delta_\text{p}(r_1,r_2) \ll 1$ for
  the all $(r_1,r_2)$ in the data set.  This is illustrated by the
  $k_\text{i} = 1000$ s$^{-1}$ case in Fig.~3A, corresponding to the
  $\Omega_r(f)$ curves in the bottom panel of Fig.~2D.
  $\Delta_\text{p}$ values close to zero translate into near total
  overlap of the $\Omega_r(f)$ results.  This regime requires
  adiabaticity, $k_\text{eq} > \bar{k}(r)$, and if there is any
  heterogeneity in the system, the interconversion between states has
  to be fast, $k_\text{i} > \bar{k}(r)$.  Thus the maximum value of
  $\bar{k}(r)$ over all $r$ gives a lower bound on both $k_\text{i}$
  and $k_\text{eq}$.

\end{enumerate}

\begin{figure*}
\centering\includegraphics[width=0.95\textwidth]{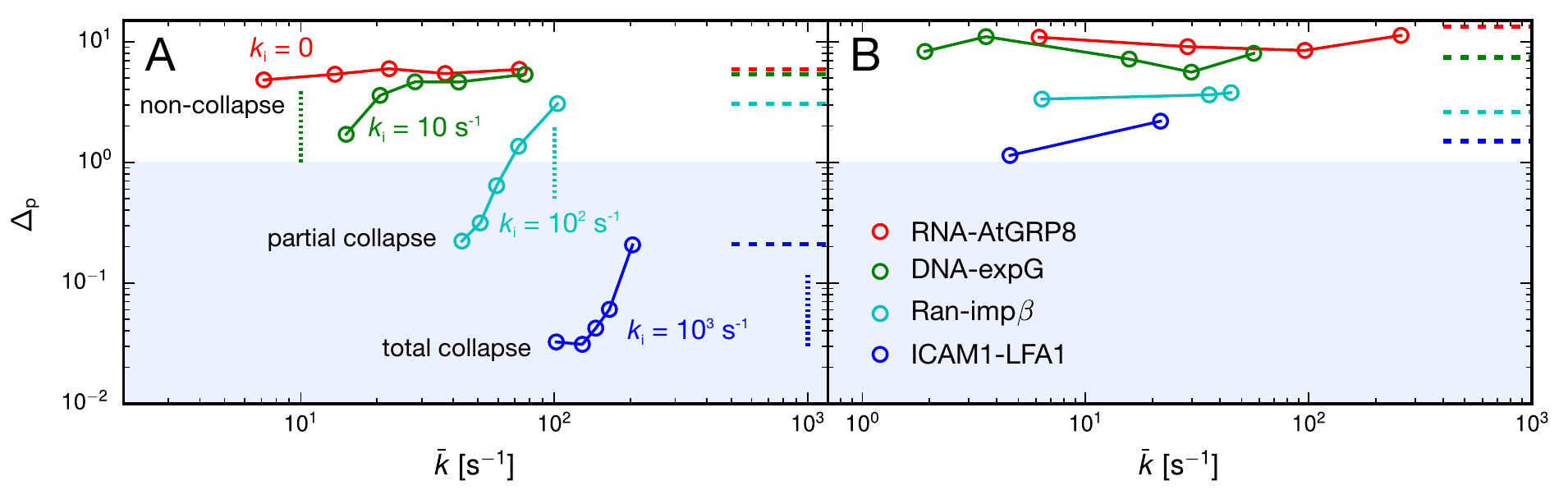}
\caption{Pair heterogeneity parameter $\Delta_\text{p}$,
  calculated from a best-fit of $\Omega_r(f)$ curves for two
  consecutive values of loading rate $(r_1,r_2)$ in a given data set.
  The horizontal axis coordinate is the smaller of the mean rupture
  rates for each pair,
  $\bar{k} = \text{min}(\bar{k}(r_1),\bar{k}(r_2))$.  For comparison,
  the $\Delta$ calculated from all loading rates in a data set is
  shown as a horizontal dashed line.  The shaded region corresponds to
  $\Delta_\text{p} \le 1$, where disorder is negligible.  A) Results
  for the FBL model system of Fig.~2, with $\sigma = 0.05$ and
  $k_\text{i} = 0$, 10, 10$^2$, and 10$^3$ s$^{-1}$.  From left to
  right, the $\Delta_\text{p}$ points for each $k_\text{i}$ value
  correspond to loading rate pairs: $(r_1,r_2) = $ (200, 500), (500, 1000), (1000,
  2000), (2000, 5000), and (5000, 10000) pN/s.  Vertical dotted lines
  mark the values of $k_\text{i}$ in each case.  Systems where
  $\Delta_\text{p} \ge 1$ across all measured time scales of
  $\bar{k}(r)$ must have slow conformational interconversion,
  $k_\text{i} < \bar{k}(r)$ or static disorder ($k_\text{i} = 0$), and
  thus correspond to the non-collapse (NC) regime.  When some
  $\bar{k}(r)$ are larger than $k_\text{i}$ and some are smaller, we
  are in the partial collapse (PC) regime, with smaller $\bar{k}(r)$
  exhibiting $\Delta_\text{p} \ll 1$, and the larger ones
  $\Delta_\text{p} \ge 1$.  When $k_\text{i} > \bar{k}(r)$ for the
  entire data set, all $\Delta_\text{p} \ll 1$, and we are in the
  total collapse (TC) regime. B) Results for four experimental systems
  (Fig.~6) that exhibit heterogeneity and have datasets with
  at least three loading rates.  The $\Delta_\text{p}$ calculated from
  pairs of loading rates are consistent with the $\Delta$ calculated
  from the total data set, and all fall in the $\Delta_\text{p} \ge 1$
  NC regime.}\label{pair}
\end{figure*}

To summarize, we can use the magnitude of the heterogeneity parameters
($\Delta$ or $\Delta_p$ depending on whether we look at the whole data
set or pairs of ramp rates) to make specific inferences about the
nature of the biomolecular free energy landscape.  $\Delta \gg 1$
(large disorder) in an experimental data set implies the following
facts: there is an ensemble of folded/intact states in the system,
these states have substantially different force-dependent rates of
rupture, and the system will only rarely switch from one state to
another before rupture occurs.  A small but finite $\Delta$ in the
range $0 \ll \Delta \lesssim 1$ (low disorder) indicates that
heterogeneity is still present, but one or both of the following are
true: the interconversion rate $k_i$ is comparable to the mean rupture
rates, so heterogeneity is partially averaged out due to transitions
between states, or the differences in rupture rate functions between
states are small.  Finding $\Delta \approx 0$ (no disorder) indicates
that either there is no heterogeneity (a single native state) or that
$k_i$ is so large that the ensemble of native states behaves
effectively like a single state.

\vspace{1em}

\begin{figure}[t]
\centering\includegraphics[width=\columnwidth]{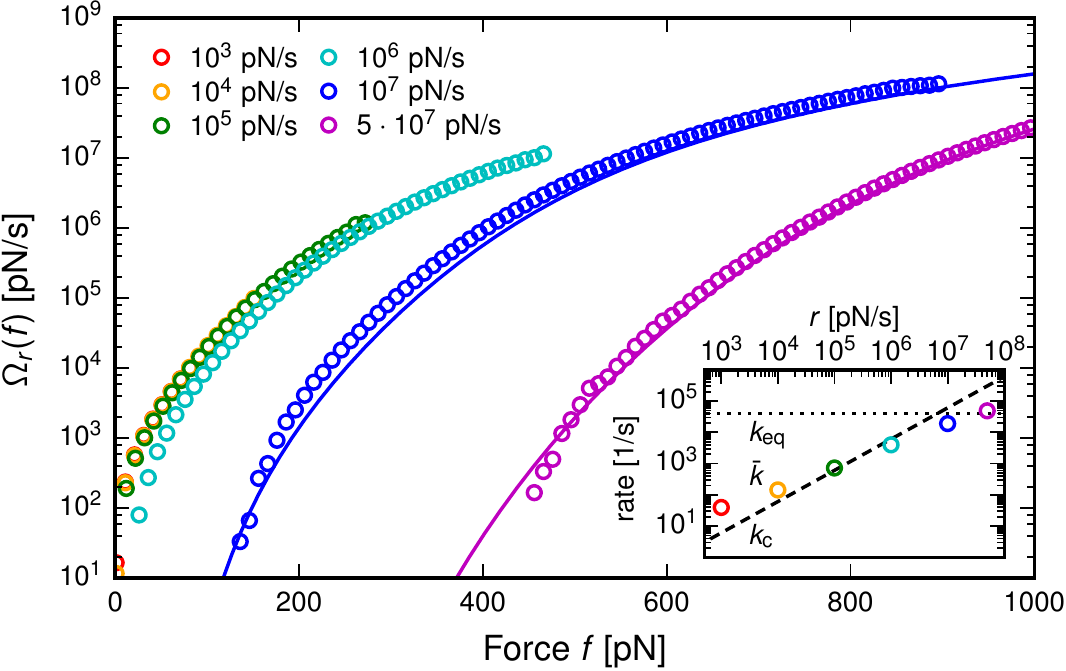}
\caption{Simulation results (circles) of {$\Omega_r(f)$} for the FBL model
  system of Fig.~2, with no disorder ($\sigma =0$) over a
  range of loading rates $r$ extending into non-adiabatic regime.
  Each color is a different value of $r$.  The solid curves for the
  two largest $r$ are plots of the analytical expression in
  Eq.~\eqref{e8}, derived for the model system in the $r\to\infty$
  limit.  The inset shows the mean rupture rate $\bar{k}(r)$ (circles)
  as a function of $r$ compared to $k_\text{eq}$ (dotted line) and
  $k_\text{c}(r)$ (dashed line).}\label{nad}
\end{figure}

{\em Ruling out non-adiabatic artifacts:} One important question about
the usefulness of the theory remains: what about situations where the
loading rate $r$ is sufficiently fast that the adiabatic assumption
$k_\text{c}(r) \ll \bar{k}(r) \ll k_\text{eq}$ breaks down?  As
mentioned above, {$\Omega_r(f)$} in this case will not collapse onto a
single master curve independent of $r$, regardless of the presence of
underlying heterogeneity in the system.  Since the experimentalist has
no direct way of measuring $k_\text{eq}$ or $k_\text{c}(r)$, it is
not {\em a priori} clear whether a given loading rate $r$ is slow
enough for adiabaticity to hold.  Can the theory in
Eqs.~\eqref{e6}-\eqref{e7} fit a pure system over a range of
non-adiabatic $r$, and yield a non-zero fitted value of $\Delta$ that
would incorrectly indicate the presence of heterogeneity?  To rule out
the possibility of such a false positive, we simulated the FBL model
system above, without any heterogeneity ($\sigma = 0$), over a much
larger range of loading rates $r$, and plotted the results of
{$\Omega_r(f)$} in Fig.~4 on a logarithmic scale for $r = 10^3 -
5\cdot 10^7$ pN/s.  As shown in the figure inset, for $r \lesssim
10^5$ pN/s, $\bar{k}(r)$ still falls between $k_\text{c}(r)$ and
$k_\text{eq}$, so adiabaticity holds and the {$\Omega_r(f)$} curves are
nearly indistinguishable.  However for $r \gtrsim 10^5$ pN/s the
collapse begins to break down, and the {$\Omega_r(f)$} curves grow
increasingly distinct.  Crucially, this non-adiabatic trend for a pure
system is qualitatively different from what happens in the adiabatic
heterogeneous case.  In the former, the curves on a logarithmic plot
grow more and more separated as $r$ grows (Fig.~4), while in the latter
situation the {$\Omega_r(f)$} curves get closer together with increasing
$r$ (Fig.~2C).  Thus a theory like Eq.~\eqref{e6}-\eqref{e7}, where convergence
at large $r$ is present ({$\Omega_r(f) \to \kappa_1(f)$} as $r$
increases), would not fit the non-adiabatic {$\Omega_r(f)$} data,
preventing a false positive.  Indeed, for the model system used in our
simulations, an expression for {$\Sigma_r(f)$} in the non-adiabatic $r
\to \infty$ limit can be analytically derived (details are in the SI) from an integral
equation approach~\cite{Hu2010},
{
\begin{equation}\label{e8}
\begin{split}
 \Sigma_r(f)\to \frac{1}{2}\left(1+\text{erf}\left[\frac{\beta D x^{\ddagger}_0 \omega_0^2 - r\left(e^{-\gamma}+\gamma-1\right)}{D\sqrt{2\beta \omega_0^3(1-e^{-2\gamma})}}\right]\right),
\end{split}
\end{equation}
}
where {$\gamma \equiv \beta D f \omega_0/r$}.  The corresponding
analytical form for {$\Omega_r(f) = -r\log\Sigma_r(f)$} is plotted in
Fig.~4 as solid curves for the two largest values of $r$,
comparing well with the simulated results.  
From Eq.~\eqref{e8} we can
explicitly see that for a fixed {$f$}, {$\Sigma_r(f) \to 1$} and
{$\Omega_r(f) \to 0$ as $r\to\infty$}, so that the {$\Omega_r(f)$} curves
on a logarithmic plot like Fig.~4 are pushed increasingly
downwards, the opposite trend of the theory in
Eqs.~\eqref{e6}-\eqref{e7}.  Thus in general, we should be able to
distinguish data sets corresponding to pure, non-adiabatic
{$\Omega_r(f)$} from heterogeneous, adiabatic ones, and false positives
can be avoided.

\vspace{1em}

{\em Analysis of experimental data:} As a demonstration of the wide
applicability of our method, we have analyzed ten
earlier datasets from biomolecular force ramp experiments, spanning a
range of scales from strand separation in DNA oligomers up to the
unbinding of large receptor-ligand complexes.  Five of
these systems (Fig.~5) showed TC of the {$\Omega_r(f)$}
curves, within experimental error bars, while the  other
five showed NC, and hence heterogeneity (Fig.~6).  Let us
consider each of these two groups in more detail.

\begin{figure*}
\includegraphics[width=\textwidth]{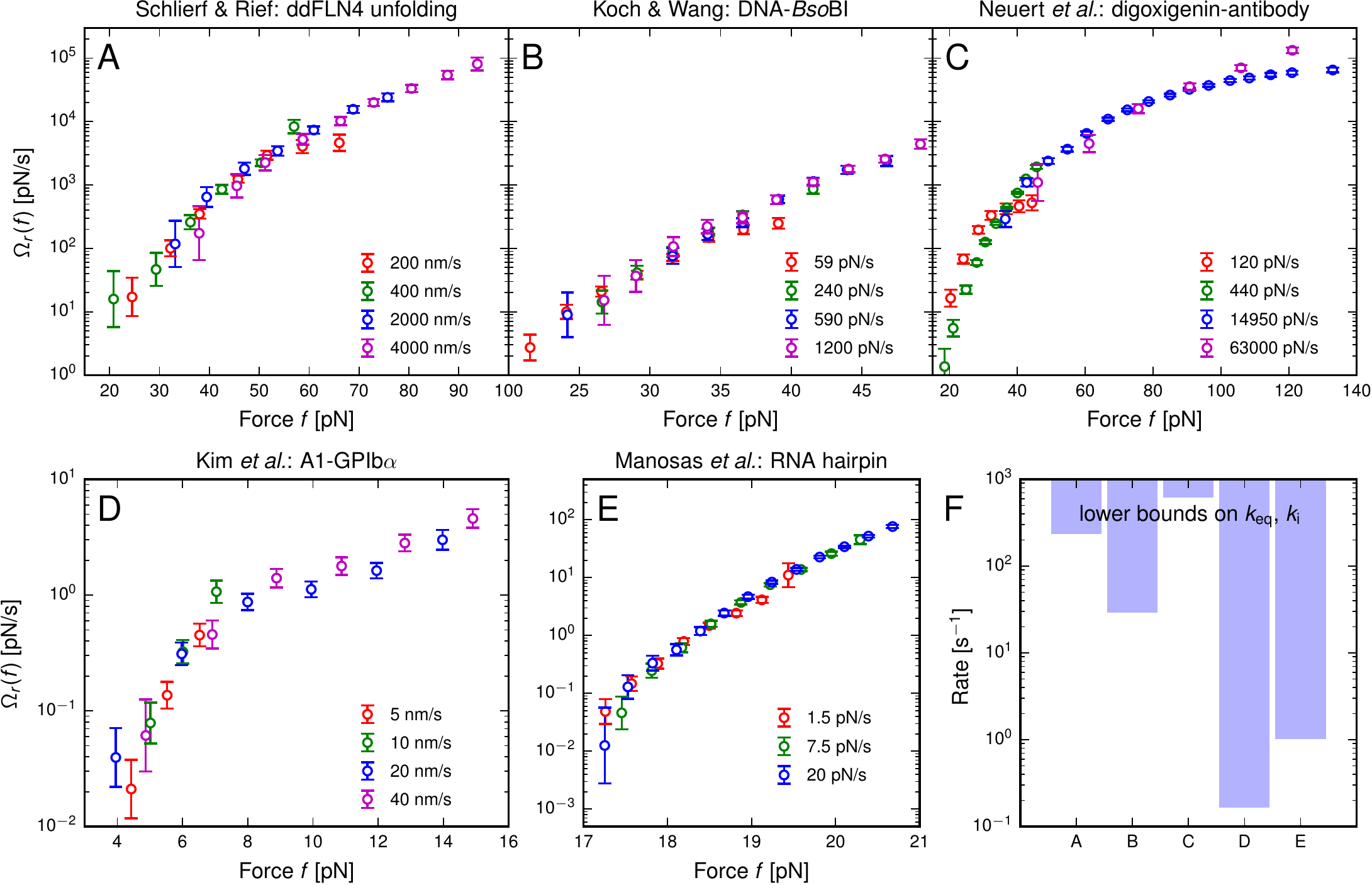}
\caption{Experimental {$\Omega_r(f)$} data (circles) calculated from
  rupture force distributions in five studies: A)
  Ref.~\cite{Schlierf2006}; B) Ref.~\cite{Koch2003}; C)
  Ref.~\cite{Neuert2006}; D) Ref.~\cite{Kim2010}.   E)
    Ref.~\cite{Manosas06}. All these cases exhibit no apparent
  heterogeneity, with the {$\Omega_r(f)$} curves for each system
  collapsing on one another.  Colors denote different pulling
  velocities $v$ or loading rates $r$, as reported in each study.  For
  A and D, where $v$ is reported, the linker stiffness values of
  {$\bar{\omega}_s = 4.1$} (A) and 0.043 pN/nm (D) are used to get the
  corresponding loading rates $r = \bar{\omega}_s v$.   F)
  For each of the experimental cases, the lower bounds on the possible
  values of $k_\text{eq}$ and $k_\text{i}$, derived from the
  theoretical analysis.}\label{col}
\end{figure*}

{\em Systems exhibiting TC:} The five experimental studies exhibiting
TC in Fig.~5 are: A) Schlierf \& Rief~\cite{Schlierf2006}, the
unfolding of immunoglobulin-like domain 4 (ddFLN4) from {\em
  D. discoideum} F-actin cross-linker filamin. B) Koch \&
Wang~\cite{Koch2003}, the unbinding of a complex between the
restriction enzyme {\em Bso}BI and DNA. C) Neuert {\em et
  al.}~\cite{Neuert2006}, the unbinding of the steroid digoxigenin
from an anti-digoxigenin antibody. D) Kim {\em et al.}~\cite{Kim2010},
the unbinding of the von Willebrand factor A1 domain from the
glycoprotein Ib $\alpha$ subunit (GPIb$\alpha$).  E) Manosas {\em et
  al.}~\cite{Manosas06}, unzipping of an RNA hairpin.  In the hairpin
case, the collapse of the $\Omega_r(f)$ curves is consistent with
collapse seen in other dynamical quantities extracted from the data at
different loading rates, for example the rupture rate $k(f)$ or the
effective barrier height at a given
force~\cite{Manosas06,Bizarro12}.  In all of the above
  experiments the data is originally gathered as time traces of the
  applied force.  The rupture or unfolding event in each trace is
  identified as a large drop in the force when using AFM (or a large
  increase in the end-to-end distance using optical tweezers), a
  signature easily detected due to its high signal-to-noise ratio.
  The value of the force immediately before the drop is then recorded.
  From hundreds of such traces, the experimentalists construct the
  distribution of forces $p_v(f)$ or $p_r(f)$ at which the system
  unfolds/ruptures for a fixed pulling velocity $v$ or loading rate
  $r$.  In those cases (A,D) where data is reported in terms of $v$
rather than $r$, mean values of the linker stiffness
{$\bar{\omega}_s$} are used to get corresponding loading rates
{$r = \bar{\omega}_s v$} (see the figure caption for values).  The
distribution {$p_r(f)$} is related to {$\Sigma_r(f)$} through
{$p_r(f) = -d\Sigma_r(f)/df$}.  By integrating {$p_r(f)$} we obtain
{$\Sigma_r(f)$} and hence {$\Omega_r(f)$}.  We can also calculate the
mean rupture force {$\bar{f}(r) = \int_0^\infty df\,f p_r(f)$} and
thus the mean rupture rate {$\bar{k}(r) = r/\bar{f}(r)$}.  The largest
value of $\bar{k}(r)$ among all the $r$ for a given experiment is
shown in the bar chart of Fig.~5F.  As mentioned above in discussing
the TC scenario, the maximum observed value of $\bar{k}(r)$ provides a
lower bound for both $k_\text{eq}$ and $k_\text{i}$.

The local equilibration rate $k_\text{eq}$ defines an intrinsic time
scale whether or not the system is heterogeneous, but the slower
interconversion rate $k_\text{i}$ exists as a distinct time scale only
when there is a heterogeneous ensemble of states with sufficiently
large energy barriers between them.  Observing collapse of
{$\Omega_r(f)$} over a range of $r$ does not absolutely rule out
heterogeneity, but it does constrain the possible values of
$k_\text{i}$.  The two systems in Fig.~5 with the strongest
constraints on $k_\text{i}$ (the largest lower bounds) are A and C,
where any $k_\text{i}$ (or $k_\text{eq}$) must be $> {\cal
  O}(10^2\:\text{s}^{-1})$.  This is not surprising, since A is a
single, compact protein domain, and C is a tight antibody complex.
For these systems, where specificity of the interactions stabilizing
the functional state is of a prime importance, significant
heterogeneity is unlikely, since it would require at least two
conformational states involving substantially different sets of
interactions.  For the more general category of enzyme-substrate or
receptor-ligand complexes (which encompasses systems B and D in
Fig.~5 and all but one of the systems in Fig.~6),
specificity may not always be the most important factor.
Conformational heterogeneity among bound complexes could play crucial
biological roles, as a part of enzymatic regulation or signaling.

System D of Fig.~5 presents an intriguing case, since force
ramp experiments on the A1-GPIb$\alpha$ complex show evidence of two
bound conformational states: a weaker bound state, from which the
system is more likely to rupture at small forces ($\lesssim 10$ pN),
and a more strongly bound state, predominating at larger
forces~\cite{Kim2010}.  The interconversion rates between the states
could not be measured, but based on fitting the ramp data to a
two-state model are estimated to be on the order of
$\sim {\cal O}(1\:\text{s}^{-1})$.  However the four experimental
pulling velocities are so slow that the mean rupture rate at the
highest velocity ($v = 40$ nm/s) is only 0.16 s$^{-1}$.  Hence, if the
two states do exist, they get averaged out over the timescale of
rupture, leading to a set of {$\Omega_r(f)$} curves that are collapsed.
We can thus make a prediction for this particular system---assuming
the two-state picture is reasonable and that both states are populated
in the ensemble of complexes at the start of the force ramp.  If the
measurements were extended to velocities significantly above 40 nm/s,
where rupture could occur on average before interconversion, the
expanded data set should exhibit PC of the {$\Omega_r(f)$} curves.  As in
the middle panel of Fig.~2D in the heterogeneous model system,
the values of $\bar{k}(r)$ where PC occurs would roughly coincide with
the interconversion rate $k_\text{i}$.  This would be one way of
directly estimating the scale of $k_\text{i}$ from experiment.

\begin{figure*}
\includegraphics[width=\textwidth]{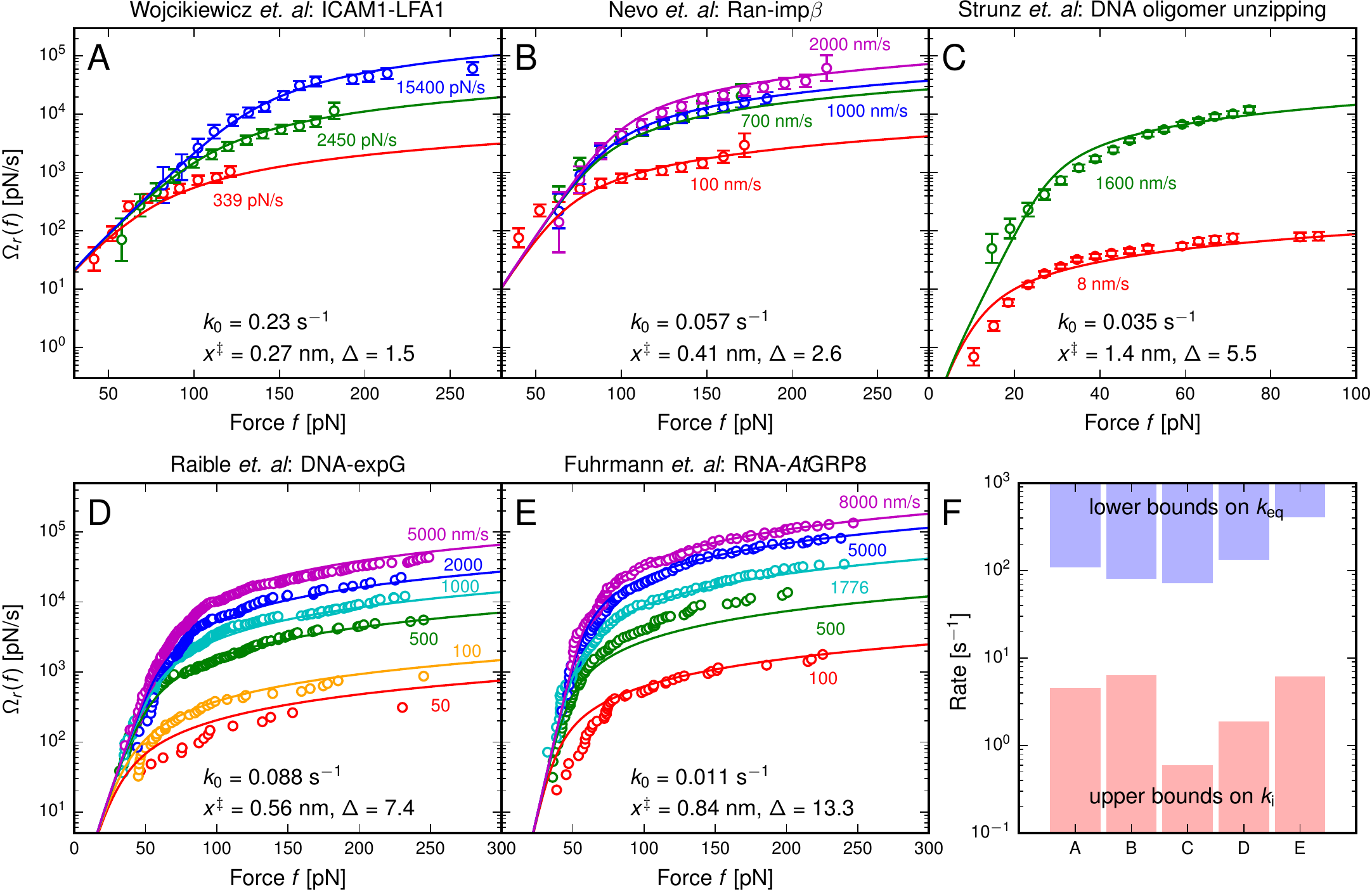}
\caption{Experimental {$\Omega_r(f)$} data (circles) calculated from
  rupture force distributions in five studies: A)
  Ref.~\cite{Wojcikiewicz2006}; B) Ref.~\cite{Nevo2004}; C)
  Ref.~\cite{Strunz1999}; D) Ref.~\cite{Raible2006BJ}; E)
  Ref.~\cite{Fuhrmann2009}.  In contrast to Fig.~5, these
  systems exhibit heterogeneity, with distinct {$\Omega_r(f)$} curves.
  Colors denote different pulling velocities $v$ or loading rates $r$,
  as reported in each study.  For B-E, where $v$ is reported, the
  linker stiffness values of {$\bar{\omega}_s = 5.0$} (B), 2.0
  (C), 3.0 (D), and 6.0 pN/nm (E) are used to get the corresponding
  loading rates {$r = \bar{\omega}_s v$}.  Solid curves show the
  theoretical best-fit to Eqs.~\eqref{e6}-\eqref{e7}, with the fitted
  parameters $k_0$, {$x^{\ddagger}$}, and $\Delta$ listed in each panel.  F) For
  each of the experimental cases, the lower bounds on the possible
  values of $k_\text{eq}$ (blue bars) and the upper bounds on
  $k_\text{i}$ (pink bars), derived from the theoretical
  analysis.}\label{ncol}
\end{figure*}

{\em Heterogeneous systems:} In contrast to Fig.~5, the five
experimental studies of Fig.~6 all show clear NC, and thus
evidence of heterogeneity: A) Unbinding of the leukocyte
function-associated antigen-1 (LFA1) integrin from its ligand,
intercellular adhesion molecule-1 (ICAM1)\cite{Wojcikiewicz2006}. B)
Rupture of the GTPase protein Ran from the nuclear receptor importin
$\beta$ (imp$\beta$)~\cite{Nevo2004}.  For this dataset, Ran is loaded
with a GTP analog (GppNHp), as well complexed with another binding
partner, the protein RanBP1.  C) Unzipping of a 10 basepair DNA
duplex~\cite{Strunz1999}.  D) Raible {\em et al.}~\cite{Raible2006BJ}
(based on earlier experimental data from Ref.~\cite{Bartels2003}), the
unbinding of the regulatory protein expG from a promoter DNA fragment;
E) Fuhrmann {\em et al.}~\cite{Fuhrmann2009}, the unbinding of the
protein {\em AT}GRP8 (in the mutant {\em AT}GRP8-RQ form) from its RNA
target.

In all these cases the theoretical fit to Eqs.~\eqref{e6}-\eqref{e7}
(solid curves) is excellent, allowing us to extract the fitting
parameters listed in each panel of Fig.~6.  The values of
$k_0$, the effective zero-force off-rate, are in the range
$\sim {\cal O}(0.01 - 0.1\:\text{s})$, while the effective transition
state distance $b \sim {\cal O}(0.1-1\:\text{nm})$.  Both of these
scales are physically sensible for protein or nucleic acid systems.
The panels in Fig.~6 are ordered by increasing $\Delta$,
which varies from 1.5 to 13.3.  To verify the robustness
  of these $\Delta$ values, we also calculated the pair parameters
  $\Delta_\text{p}$ for every data set that had at least three different
  loading rates.  These are shown in Fig.~3B, with the
  corresponding $\Delta$ for the full data indicated as horizontal
  dashed lines.  As is expected for the NC regime, the $\Delta_\text{p}$ do
  not vary significantly with rupture rate, and are consistent with
  $\Delta$ in each case.  The three largest values of $\Delta$
(Fig.~6C-E) correspond to bonds composed of nucleic acid
base-pairing or protein/nucleic acid interactions.  This significant
heterogeneity may reflect the tendency for free energy landscapes
involving nucleic acids to be more intrinsically rugged.  However it
is not necessarily the case that all nucleic acid systems are
heterogeneous (the {\em Bso}BI-DNA complex of Fig.~5B and the RNA hairpin of Fig.~5E are counter-examples).

All the data in Fig.~6 was collected using AFM pulling experiments, in
contrast to Fig.~5, where panels B, D, and E were optical trap results
(the rest being AFM). It is thus worthwhile to wonder whether aspects
of the AFM experimental setup could affect the heterogeneity analysis.
In SI Sec.~5 we have analyzed possible errors from several sources:
the finite force resolution of AFM cantilever, the non-negligible
hydrodynamic drag on the cantilever at large pulling speeds (> 1
$\mu$m/s)~\cite{Alcaraz2002,Janovjak2005,Liu2010}, uncertainties
arising from finite sampling of the rupture force distributions, and
the apparatus response time.  Based on this error analysis, we
conclude that the estimation of the heterogeneity parameter $\Delta$
from the experimental data is reliable in all the systems of
Fig.~6. The observed heterogeneity must therefore be an intrinsic
aspect to the biomolecules, rather than an artifact of the AFM
experiment.

The fidelity of the theoretical fits to the data in Fig.~6
(with no signs of PC) means all the experiments were in
the heterogeneous, adiabatic regime.  Thus the range of observed
$\bar{k}(r)$ allows us to place upper bounds on $k_\text{i}$ and lower
bounds on $k_\text{eq}$, which are plotted in the bar chart of
Fig.~6F.  There is a clear separation of time scales, with
all the upper bounds on $k_\text{i} \lesssim 10$ s$^{-1}$, and the
lower bounds on $k_\text{eq} \gtrsim 10^2$ s$^{-1}$.  The slow
interconversion rates $k_\text{i}$ in these systems are remarkable,
particularly the DNA oligomer in Fig.~6C, which is a tiny
system only 10 basepairs long.  The rupture force distributions for
the DNA unzipping were earlier fit to a specific model of dynamic
disorder in Ref.~\cite{Hyeon2014}, where force-dependent rates of
conformational fluctuations were extracted.  The range of these
estimated rates ($2.8\times 10^{-5} - 4.8\times 10^{-1}$ s$^{-1}$) are
consistent with the upper bound derived from the current analysis,
$k_\text{i} < 0.6$ s$^{-1}$.  However, we must keep in mind
that---unless PC is observed, pinpointing the scale of
$k_\text{i}$---our analysis cannot distinguish between a heterogeneous
system characterized by dynamic disorder with slow $k_\text{i}$ and
one with quenched disorder ($k_\text{i}=0$) caused by covalent chemical
differences among the experimental samples.

The Ran-imp$\beta$ system in Fig.~6B provides an interesting
counterpart to the A1-GPIb$\alpha$ complex discussed earlier.  As in
that example, the system is believed to exhibit two bound
conformations with different adhesion
strengths~\cite{Nevo2003,Nevo2004}.  This is also supported by
evidence of conformational variability in the crystal structure of a
truncated imp$\beta$ bound to Ran-GppNHp, where two versions of the
molecular complex were observed, characterized by substantially
different sets of interactions~\cite{Vetter1999}.  The bound
conformations are expected to dynamically interconvert, but the
timescale has not been measured.  Our analysis of the existing data
provides an upper bound on the rate, $k_\text{i} < 6.4$ s$^{-1}$.  We
predict that further experiments could fix the rate more precisely:
for example, by going to pulling velocities slower than $v = 100$ nm/s
(the slowest $v$ in the current dataset), we may be able to observe
PC, like in the middle panel in Fig.~2D, establishing the scale of $k_\text{i}$.  This is
opposite of the prescription we gave above for the A1-GPIb$\alpha$
complex, where the existing experiments have been too slow rather than
too fast.  Our theory thus provides a guide for experimentalists to
fine-tune their parameters to extract the most information possible
from the system under study.

We envision that our approach will become one part of a larger,
comprehensive experimental toolbox for investigating heterogeneity in
biomolecules: it can test for and quantify heterogeneity based on the
rupture force distributions, but these distributions do not contain
all the information we would like to know about a system.  A large
$\Delta$ parameter indicates that there are multiple states in the
intact/folded part of the free energy landscape, and that these states
must interconvert on timescales slower than the mean rupture time.  To
extract additional details, like the precise number of functional
states, requires using other experimental/analytical techniques, like
single-molecule FRET.  One recent example where this was demonstrated
was the $k$-means clustering algorithm applied by Hyeon {\it et
  al.}~\cite{Hyeon2012} to estimate the number of interconverting
states from single-molecule FRET trajectories of a simple nucleic acid
construct, the Holliday junction. In principle, this approach could be
extended to folding trajectories obtained in constant force
experiments, which in conjunction with the distribution of rupture
forces could be used to extract the number of distinct functional
states.

\section*{Conclusions}

Our work introduces a generic method for characterizing heterogeneity
in biomolecules using rupture force distributions from force
spectroscopy experiments.  The central result is a single
non-dimensional parameter $\Delta \ge 0$.  A system with no measurable
heterogeneity on the timescale of the pulling experiment has
$\Delta = 0$.  When $\Delta > 0$, its magnitude characterizes the
degree of the disorder.  Both in the presence and absence of
heterogeneity, the method yields bounds on the local equilibration
rate $k_\text{eq}$ within a system state, and (if heterogeneity is
present) the rate of interconversion $k_\text{i}$ between states.  The
practical value of our approach is demonstrated by analyzing nine
previous experiments, allowing us to classify a broad range of
biomolecular systems.  The five cases where heterogeneity was observed
are all the more striking given the persistence of their
conformational states, with upper bounds on $k_\text{i} \lesssim 10$
s$^{-1}$.  

Our theory leads to a proposal for future experimental studies:
searching for a range of pulling speeds where the data exhibits the
property of partial collapse, allowing for a more accurate
determination of $k_\text{i}$.  This PC scenario did not occur among
the data sets we considered, though in two cases (the protein
complexes A1-GPIb$\alpha$ and Ran-imp$\beta$) we predict that
extending the range of pulling velocities would very likely result in
PC.  The global energy landscapes of multi-domain protein and nucleic
acid systems are essential guides to their biological function, but
are quite difficult to map out in the laboratory.  This is
particularly true for systems where the ruggedness of the landscape
creates a host of long-lived, functional states.  The theory described
here suggests new ways in which single molecule pulling experiments
can be used to obtain information about internal dynamics of systems
with functionally heterogeneous states.  Our technique should shed new
light on both the static and dynamic aspects of such landscapes, the
first step towards a comprehensive structural understanding of these
biomolecular shape-shifters.

\begin{acknowledgments}
This work was initiated when M.H. and D.T. were visiting scholars in KIAS in 2013.
\end{acknowledgments}

\newpage

\setcounter{equation}{0}
\setcounter{table}{0}
\setcounter{figure}{0}

\renewcommand{\theequation}{S\arabic{equation}}
\renewcommand{\thetable}{S\arabic{table}}
\renewcommand{\thefigure}{S\arabic{figure}}
\renewcommand\thesection{\arabic{section}}

\onecolumngrid

\begin{center}
{\Large Supplementary information for:\\[0.5em]  ``Directly measuring single molecule heterogeneity using force spectroscopy''}\\[1em]

\large Michael Hinczewski, Changbong Hyeon, D. Thirumalai
\end{center}

\section{Testing the assumptions of the $\Omega_r(f)$ model with respect to possible generalizations}\label{test}

The general form for $\Omega_r(f)$ introduced in Eq.~(6) of the main
text,
\begin{equation}\label{t1}
\Omega_r(f) \approx \frac{r}{\Delta(f)} \log \left(1+\frac{\kappa_1(f)\Delta(f)}{r}\right),
\end{equation}
depends on the functions $\Delta(f)$ and $\kappa_1(f)$.  For the
analysis of the experimental data, we chose a minimal model for these
two functions, shown in Eq.~(7):
\begin{equation}\label{t2}
\Delta(f) = \Delta, \qquad \kappa_1(f) = \frac{k_0}{\beta x^{\ddagger}} \left(e^{\beta f x^{\ddagger}}-1\right).
\end{equation}
This assumes $\Delta(f)$ is constant across the force range of the
experiment, and $\kappa_1(f)$ takes the same mathematical form as in
the case of a pure Bell model,
$\kappa_1(f) = \int_0^f df^\prime k(f^\prime)$, where
$k(f) = k_0 e^{\beta f x^\ddagger}$.  The result is a three parameter
model (depending on $\Delta$, $k_0$, $x^{\ddagger}$) that is able to
simultaneously fit $\Omega_r(f)$ data for loading rates across two to
three orders of magnitude for a large number of unrelated
  biological systems.

However, it is worthwhile to ask if the general conclusions that we
draw from the experimental fitting would change substantially if the
above assumptions were relaxed, and we used more complicated forms for
$\Delta(f)$ and $\kappa_1(f)$.  Here we will examine  two
generalizations of the minimal model (in each adding another fitting
parameter) and verify that our characterization of heterogeneity in
the experimental systems is indeed robust.

\vspace{1em}

\noindent i) {\it Dudko-Hummer-Szabo model for $k(f)$:} The most widely used
generalization of the Bell model was introduced by Dudko, Hummer, and
Szabo (DHS)~\cite{Dudko2006}.  In this approach, the escape rate $k(f)$
is calculated from Kramers theory for particular choices of the
underlying 1D free energy profile, leading to
\begin{equation}\label{t3}
k_\text{DHS}(f) = k_0 \left(1-\frac{\nu f x^\ddagger}{G^\ddagger}\right)^{\frac{1}{\nu}-1} e^{\beta G^\ddagger \left[1-(1-\nu f x^\ddagger/G^\ddagger)^{1/\nu}\right]},
\end{equation}
which introduces two extra parameters: $G^\ddagger$, the height of the
free energy barrier at zero force, and $\nu$, characterizing the shape
of the 1D free energy profile, in addition to $x^\ddagger$ (the
transition state distance) and $k_0$ (the rate at zero force).  The
constant $\nu$ is usually chosen to be either 2/3 or 1/2,
corresponding to linear-cubic or cusp-like free energy profiles
respectively.  We use $\nu=2/3$ in the analysis below, though the
results were similar for $\nu=1/2$.  With the escape rate
$k_\text{DHS}(f)$, the generalized form for
$\kappa_1(f) = \int_0^f df^\prime k_\text{DHS}(f^\prime)$ becomes:
\begin{equation}\label{t4}
\kappa_1(f) = \frac{k_0}{\beta x^\ddagger}\left(e^{\beta G^\ddagger \left[1-(1-\nu f x^\ddagger/G^\ddagger)^{1/\nu}\right]}-1 \right).
\end{equation}
Substituting this for $\kappa_1(f)$, with $\nu$ fixed at 2/3, in Eq.~\eqref{t2} we have a four
parameter model for $\Omega_r(f)$, depending on $\Delta$, $k_0$,
$x^{\ddagger}$, and $G^\ddagger$.

\begin{figure}[t]
\includegraphics[width=0.5\textwidth]{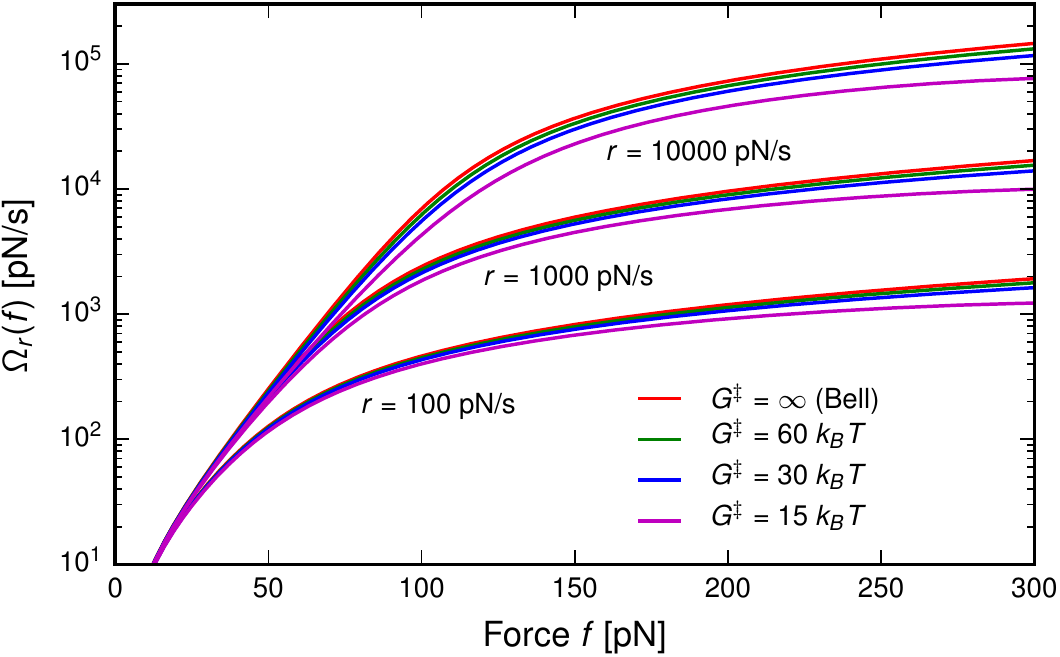}
\caption{$\Omega_r(f)$ curves of the FBL model for $r=100$, $1000$ and $1000$ pN/s
  using the DHS form for $\kappa_1(f)$ [Eq.~\eqref{t4}].  The
  parameters are $\Delta = 5$, $k_0 = 0.1$ s$^{-1}$,
  $x^\ddagger = 0.3$ nm, and various values for $G^\ddagger$, ranging from
  $\infty$ (red curves, corresponding to the Bell limit) down to 15
  $k_B T$ (purple curves).}\label{dhst}
\end{figure}

\begin{figure}
\includegraphics[width=\textwidth]{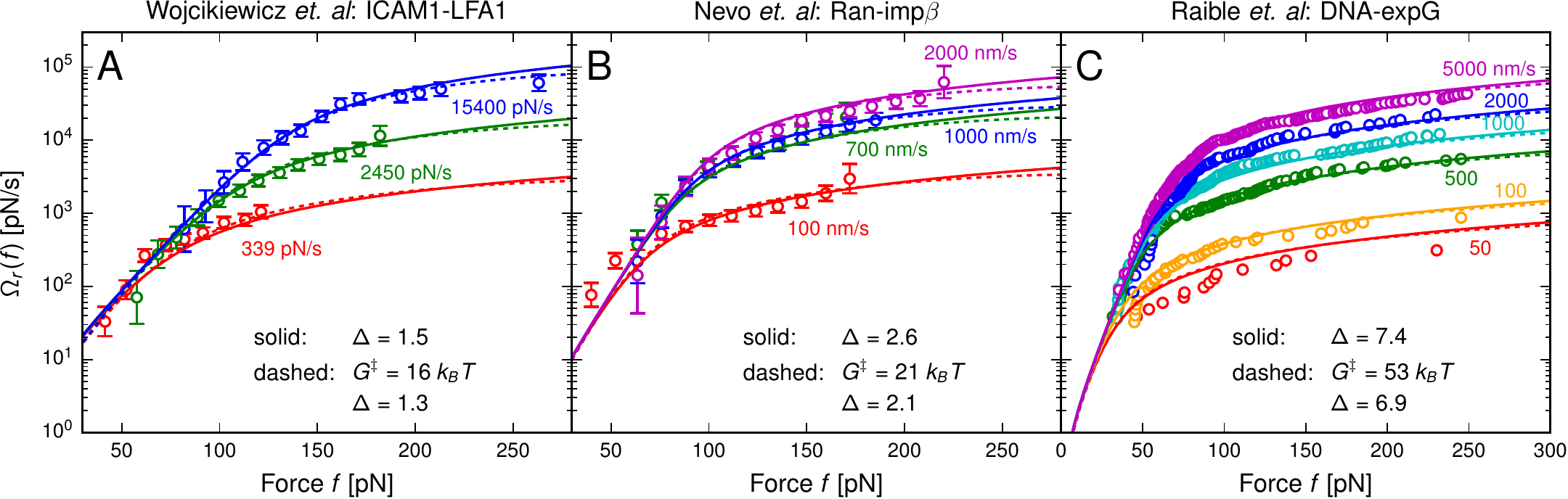}
\caption{Experimental {$\Omega_r(f)$} data (circles) calculated from
  rupture force distributions in three studies: A)
  Ref.~\cite{Wojcikiewicz2006}; B) Ref.~\cite{Nevo2004}; C)
  Ref.~\cite{Raible2006BJ}.  Colors denote different pulling
  velocities $v$ or loading rates $r$, as reported in each study.  For
  B-C, where $v$ is reported, the linker stiffness values of
  {$\bar{k}_s = 5.0$} (B) and 3.0 (C) are used to get the
  corresponding loading rates {$r = \bar{k}_s v$}.  Solid curves show
  the theoretical best-fit for the minimal three-parameter model
  (Eqs.~\eqref{t1}-\eqref{t2}) with the extracted value of $\Delta$
  indicated in the panel.  The dashed curves are the
  best-fits from the four-parameter DHS model ($\kappa_1(f)$ replaced
  by Eq.~\eqref{t4}) with the values for $G^\ddagger$ and $\Delta$
  listed at the bottom of the panel.}\label{dhs}
\end{figure}

In the limit of $G^\ddagger \to \infty$ the DHS model reduces to the
original Bell form, and hence the results for $\Omega_r(f)$ are the
same as in the minimal model.  For $G^\ddagger < \infty$ the DHS form
introduces small corrections, shown in Fig.~\ref{dhst}, particularly
at larger forces where the increase in $\Omega_r(f)$ is not as rapid
as in the Bell version.  Note that in fitting to experimental data,
the parameter $G^\ddagger$ cannot be made smaller than
$f_\text{max} \nu x^\ddagger$, where $f_\text{max}$ is the largest
force value that appears in the data set.  The DHS model is not
mathematically defined for $G^\ddagger$ below that cutoff.
Fig.~\ref{dhs} shows three sets of experimental results for
$\Omega_r(f)$ from Fig.~6 of the main text, comparing the minimal
model fits (solid curves) to the best-fit using the more complex DHS
model (dashed curves).  These three systems yielded $G^\ddagger$
values in the range $16 - 53$ $k_BT$.  (The other two experimental
systems from Fig.~6 did not exhibit any improved fitting using the DHS
model, since the best-fit $G^\ddagger$ was large enough that the
results were numerically indistinguishable from the minimal Bell
model.)  The DHS fits for $\Omega_r(f)$ in Fig.~\ref{dhs} are very
close to the minimal model fits, and the extracted $\Delta$ values
from the two approaches differ by only $5 - 20\%$, a discrepancy
comparable to the uncertainty in $\Delta$ due to finite sampling of
the rupture force distribution (see SI Sec.~\ref{exp}.ii below).  Thus, at
least for the data sets we have looked at, the Bell approximation is
justifiable, and does not affect our heterogeneity analysis in terms
of $\Delta$ in a significant way.

\vspace{1em}

\begin{figure}
\includegraphics[width=\textwidth]{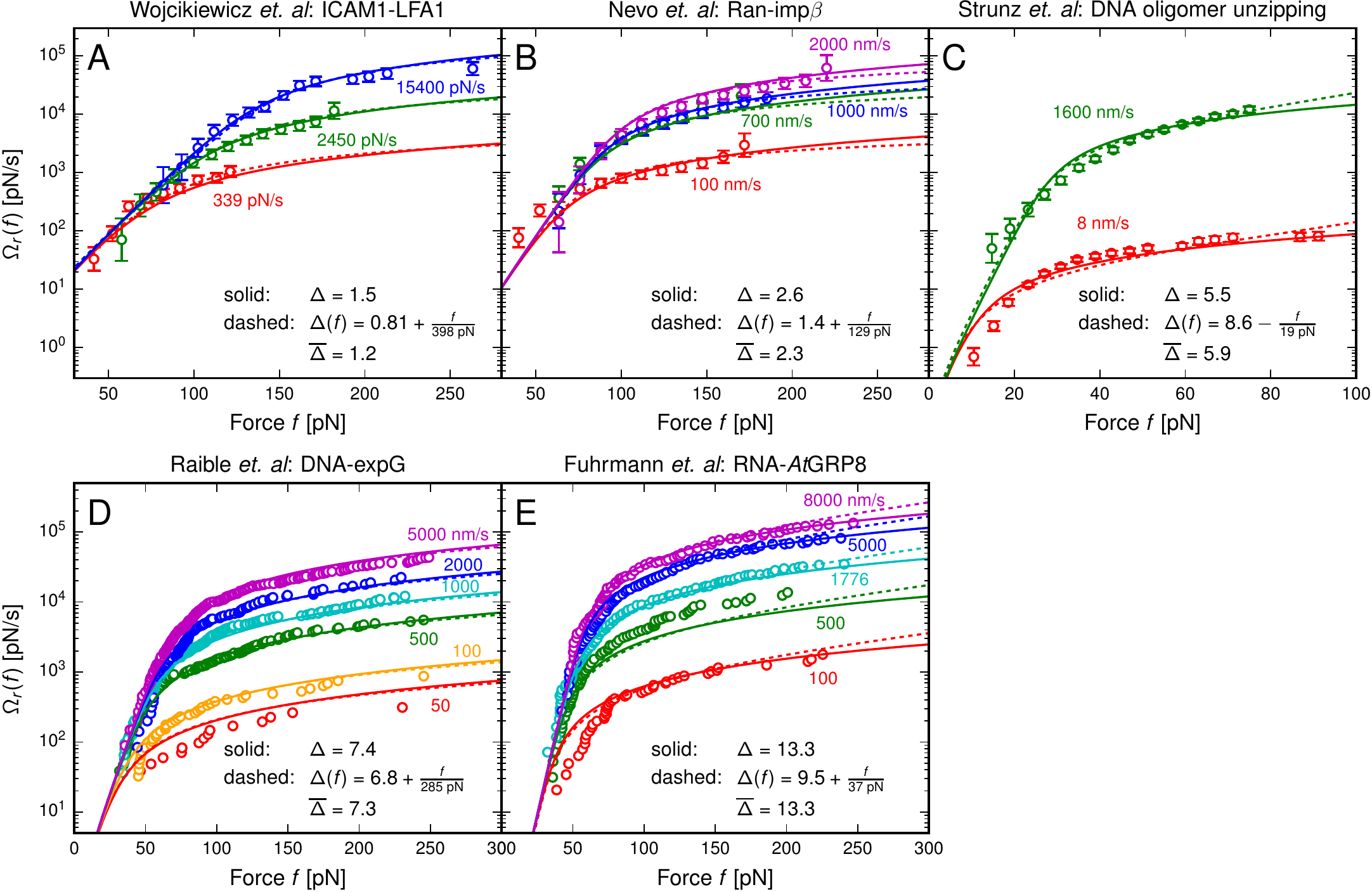}
\caption{Experimental {$\Omega_r(f)$} data (circles) calculated from
  rupture force distributions in five studies: A)
  Ref.~\cite{Wojcikiewicz2006}; B) Ref.~\cite{Nevo2004}; C)
  Ref.~\cite{Strunz1999}; D) Ref.~\cite{Raible2006BJ}; E)
  Ref.~\cite{Fuhrmann2009}.  Colors denote different pulling
  velocities $v$ or loading rates $r$, as reported in each study.  For
  B-E, where $v$ is reported, the linker stiffness values of
  {$\bar{k}_s = 5.0$} (B), 2.0 (C), 3.0 (D), and 6.0 pN/nm (E) are
  used to get the corresponding loading rates {$r = \bar{k}_s v$}.
  Solid curves show the theoretical best-fit for the minimal
  three-parameter model (Eqs.~\eqref{t1}-\eqref{t2}) with the
  extracted value of $\Delta$ indicated in the panel.  The dashed
  curves are the best-fits from the four-parameter model where
  $\Delta(f)$ varies linearly with $f$.  The fit results for
  $\Delta(f)$ are shown at the bottom of the panel, together with the
  average value $\overline{\Delta}$ of $\Delta(f)$ across the
  experimental force range.}\label{lind}
\end{figure}

\noindent ii) {\it Linearly varying $\Delta(f)$:} The second
generalization of the minimal model which we consider is relaxing the
assumption that $\Delta(f)$ is constant across the measured force
range.  At lowest order we can allow $\Delta(f)$ to be a linear
function of $f$, $\Delta(f) = \Delta_0 + f/f_0$, where $\Delta_0$ and
$f_0$ are constants.  This leads to a four parameter model for
$\Omega_r(f)$, depending on $\Delta_0$, $f_0$, $k_0$, and
$x^\ddagger$.  Fig.~\ref{lind} shows $\Omega_r(f)$ for all five
experimental systems from Fig.~6 of the main text, and compares the
best-fit results for the constant versus linear $\Delta(f)$ models.
The shapes of the $\Omega_r(f)$ curves from the two approaches are
very similar.  To compare the predicted heterogeneity from the two
models, we calculated the average $\overline{\Delta}$ of the linear
$\Delta(f)$ best-fit function across the range of experimentally
measured forces in each case.  The difference between
$\overline{\Delta}$ and the best-fit value for $\Delta$ in the minimal
model was less than 20\% in all the systems.  This confirms that
assuming constant $\Delta(f)$ in the minimal model gives a reasonable
estimate of the average of $\Delta(f)$ over the experimental force
range.  

\vspace{1em}

Thus, both generalizations of the minimal model lead to quantitatively
similar results for heterogeneity in the experimental data.  Following
the Occam's razor principle, we thus have confined our analysis in
the main text to the three parameter model for $\Omega_r(f)$, which
has the added benefit of a simpler interpretation.  However, it is
conceivable that future data sets might require one or both of these
extensions for reasonable fitting, due to specific details of the
biological system.  The generality of Eq.~\eqref{t1} easily
accommodates these extensions and more, allowing us to incorporate
complex parametrizations of $\Delta(f)$ and $\kappa_1(f)$ if
necessary.

\section{Heterogeneous model simulation details}\label{sim}

The heterogeneous model in the main text describes diffusion along a
reaction coordinate $x$ characterized by a diffusivity $D$ and a free
energy at zero force $U(x) = (1/2) \omega_0 x^2$.  If the system
undergoes pulling at a constant force ramp rate $r$, the potential
becomes time-dependent, $U(x,t) = (1/2) \omega_0 x^2 - r t x$.  Each
simulation trajectory is generated using Brownian
dynamics~\cite{Ermak78} on this potential, with parameters
$r=200-10000$ pN/s, $D = 100$ nm$^2$/s, $\omega_0 = 400$
$k_BT/$nm$^2$, $x^{\ddagger}_0=0.2$ nm.  The simulation time step is
$\Delta t = 0.1$ $\mu$s.  The system is initialized at $x=0$ and
$x^{\ddagger}=x^{\ddagger}_0$, and run until the rupture occurs, $x \ge x^{\ddagger}$. At every time
step, along with the Brownian dynamics update of $x$, we also include
the possibility of conformational interconversion as a Poisson
process: a random number $\eta$ between 0 and 1 is chosen; if
$\eta > \exp(-k_\text{i} \Delta t)$, a new value of $x^{\ddagger}$ is drawn from
the Gaussian distribution
$P(x^{\ddagger}) = \exp(-(x^{\ddagger}-x^{\ddagger}_0)^2/2\sigma^2)/\sqrt{2\pi \sigma^2}$.  The ranges
of distribution widths and interconversion rates are $\sigma = 0-0.05$
nm, $k_\text{i} = 0 - 10^4$ s$^{-1}$.  The rupture event at the end of
a trajectory occurs at a particular time $t$, corresponding to a force
$f = rt$.  By collecting about $3\times 10^4$ trajectories for each
value of $r$, we get a rupture force distribution $p_r(f)$.  The
survival probability $\Sigma_r(f)$ is the cumulative distribution
$\Sigma_r(f) = 1- \int_0^f df^\prime\, p(f^\prime)$, from which we can
then calculate $\Omega_r(f) = -r \log \Sigma_r(f)$.

\section{Heterogeneity in rupture pathways versus heterogeneity in functional states}\label{path}

\begin{figure}
\includegraphics[width=\textwidth]{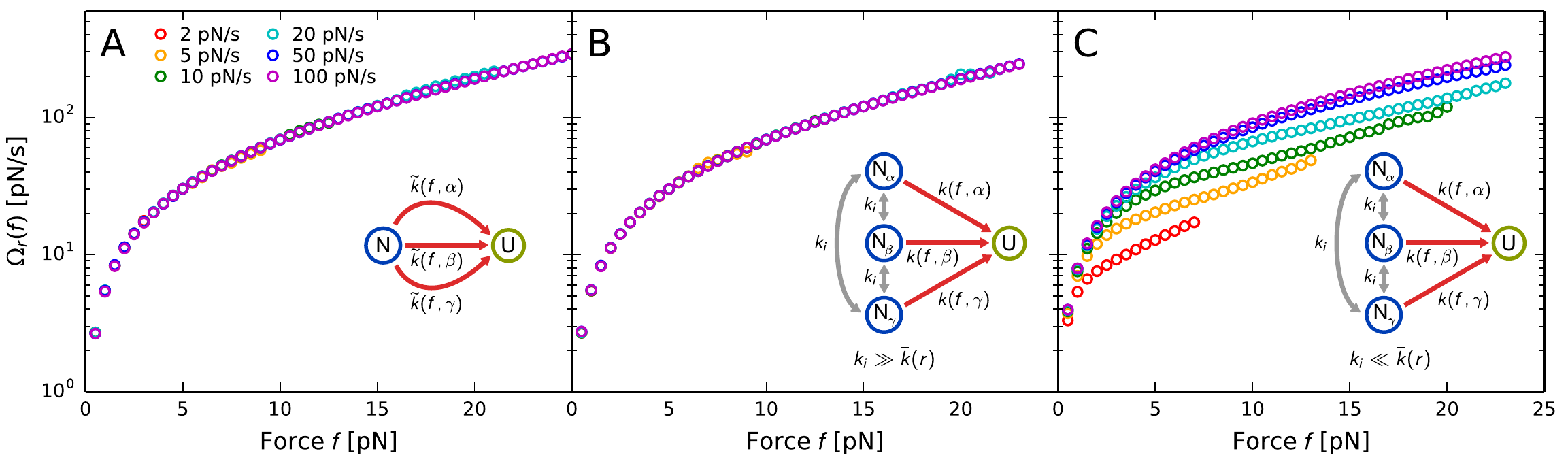}
\caption{A) A system with three different rupture pathways between the
  native (N) and unbound (U) states.  The transition rates for each
  pathway at force $f$ are given by:
  $\tilde{k}(f,\alpha) = 10\:\text{s}^{-1} e^{\beta f (0.1
    \:\text{nm})}$,
  $\tilde{k}(f,\beta) = 1\:\text{s}^{-1} e^{\beta f
    (0.5\:\text{nm})}$,
  $\tilde{k}(f,\gamma) = 5\:\text{s}^{-1} e^{\beta f
    (0.3\:\text{nm})}$.  Assuming a force ramp $f(t) = rt$, the
  corresponding master equation for the time evolution of the system
  is solved numerically, and the results for $\Omega_r(f)$ are plotted
  for six different ramp rates $r$ between $2$ and $100$ pN/s.  B)
  Analogous to panel A, but for a system with multiple native states,
  interconverting at rate $k_i$.  The rupture rate functions are a
  factor of 3 times larger than their counterparts in A, for example
  $k(f,\alpha) =3 \tilde{k}(f,\alpha)$.  The value $k_i=1000$ s$^{-1}$
  is large enough that $k_i \gg \bar{k}(r)$, the mean rate of rupture
  at each $r$.  C) Same as in panel B, but with $k_i = 0.1$
  s$^{-1} \ll \bar{k}(r)$.}\label{kmc}
\end{figure}

The heterogeneity discussed in the main text refers to the presence of
multiple, distinct functional states N$_\alpha$, each characterized by
a certain rupture rate at constant force, $k(f,\alpha)$.  But
biomolecules can also exhibit another kind of heterogeneity, where a
native basin of attraction has multiple dynamic pathways by which the
system can unfold or rupture to reach state U.  In fact, the two kinds
of heterogeneity can in principle exist in the same system.
Fig.~\ref{kmc}A depicts a simple model that is heterogeneous in the
second sense (though not the first): a single native state N has three
rupture pathways, labeled $\alpha$, $\beta$, and $\gamma$, with
corresponding rate functions $\tilde{k}(f,\alpha)$,
$\tilde{k}(f,\beta)$, and $\tilde{k}(f,\gamma)$.  At a certain force
$f$, this is equivalent to a total rate of transitioning from N to U
given by
$k(f) = \tilde{k}(f,\alpha) + \tilde{k}(f,\beta) +
\tilde{k}(f,\alpha)$.  Assuming adiabaticity under a force ramp
$f(t)$, the survival probability $\Sigma_r(t)$ obeys the kinetic
equation $d\Sigma_r(t)/dt = -k(f(t))\Sigma_r(t)$, and the same
arguments apply as for the pure system in the main text, leading to
collapse of the $\Omega_r(f)$ curves.  We verify this numerically for
the model system, solving the associated master equation.  We show the
$\Omega_r(f)$ results in Fig.~\ref{kmc}A for particular choices of
$\tilde{k}$ described in the caption.

It is instructive to compare this multiple-pathway,
single-native-state system to the functionally heterogeneous system
shown in Fig.~\ref{kmc}B.  Here there are three native states
N$_\alpha$, N$_\beta$, and N$_\gamma$, with corresponding rupture rate
functions $k(f,\alpha)$, $k(f,\beta)$, and $k(f,\gamma)$.  In the
limit $k_i \gg \bar{k}(r)$, where the rate of interconversion between
the states is much faster than the rate of transitioning to U, the
ensemble of native states gets averaged out, acting as effectively a
single state with net rupture rate
$k(f) = p_\alpha k(f,\alpha) + p_\beta k(f,\beta) + p_\gamma
k(f,\gamma)$.  Here $p_\alpha$ is the stationary probability of the
system being in state $\alpha$.  Since the interconversion rate is
identical between all pairs of states,
$p_\alpha = p_\beta = p_\gamma = 1/3$.  If we choose rate functions
such that $k(f,\alpha)= 3\tilde{k}(f,\alpha)$, and similarly for
$\beta$ and $\gamma$, we should find that the $\Omega_r(f)$ curves
collapse to the same result as in the first, multiple pathway system.  This
is indeed what the numerical results in Fig.~\ref{kmc}B show.

In contrast, if $k_i \ll \bar{k}(r)$, functional heterogeneity will
manifest itself in the $\Omega_r(f)$ curves, and we get the non-collapse
of Fig.~\ref{kmc}C.  Thus non-collapse is a signature of a particular
kind of heterogeneity: multiple native states with slow rates of
interconversion between them (i.e. due to high barriers separating the
states).  Such a system will by definition have many pathways to
rupture (at least one from each native state), but the existence of
multiple pathways is not by itself sufficient to trigger non-collapse.

The argument above has interesting implications for analyzing
distributions of forces at which biomolecules fold (rather than
unfold/rupture).  These correspond to transitions starting in the
unfolded ensemble, which should usually behave as a single state,
having sufficiently fast interconversion times due to small energy
barriers between unfolded configurations.  Even if there were multiple
pathways to fold to a single (or many) native states, the
$\Omega_r(f)$ calculated from the survival probability $\Sigma_r(f)$
of the unfolded state should exhibit collapse, assuming adiabaticity
holds.

However, it is conceivable that the unfolded state ensemble in certain
cases could be heterogeneous, partitioning into multiple states that
do not interconvert readily. In this scenario, the refolding force
distributions when analyzed using our theory would manifest
heterogeneity. If this were the case then our framework offers an
ideal way of investigating the nature of unbound complexes or unfolded
states of proteins and RNA. These issues await future experiments.

\section{Derivation of non-adiabatic survival probability}\label{nonad}

The derivation of Eq.~(9) in the main text, the non-adiabatic limit of
the survival probability $\Sigma_r(f)$ for the heterogeneous model,
follows from an approach outlined by Hu, Cheng, and Berne~\cite{Hu2010}.  This is closely related to the renewal method for
calculating first-passage time distributions~\cite{Kampen}.  We are
interested in $\Sigma_r(t)$, the probability that the system has never
reached $x=x^{\ddagger}>0 $ at time $t$, given the initial condition $x=0$ at
$t=0$.  This yields $\Sigma_r(f)$ after the change of variables from
$t$ to $f(t) = rt$.  In the  model accounting for heterogeneity described in the main text, the value of $b$ changes
randomly with an interconversion rate $k_\text{i}$.  However here we
focus only on the case with no disorder, where $x^{\ddagger}$ is fixed at a value of
$x^{\ddagger}_0$.

The survival probability $\Sigma_r(t)$ can be expressed as an integral
\begin{equation}\label{si1}
\Sigma_r(t) = \int_{-\infty}^{x^{\ddagger}_0} dx\,P(x,t),
\end{equation}
where $P(x,t)$ is the probability that the particle is at $x$ at time
$t$, having never reached $x=x^{\ddagger}_0$ at any time prior to $t$.  The
initial condition is $P(x,0) = \delta(x)$.  Because of the $x=x^{\ddagger}_0$
condition, $P(x,t)$ is difficult to calculate directly, but it is
related to the simpler Green's function $G(x,t |x^\prime, t^\prime)$
defined in the absence of any condition.
$G(x,t | x^\prime, t^\prime)$ is just the probability of being at $x$
at time $t$, given that it was at $x^\prime$ at time $t^\prime \le t$,
and assuming the particle is allowed to diffuse in the
$U(x,t) = (1/2) \omega_0 x^2 - r t x$ potential across the entire
range $-\infty < x < \infty$.  It satisfies the Fokker-Planck equation
\begin{equation}\label{si2}
\frac{\partial G}{\partial t}=D \frac{\partial}{\partial x} \left[ e^{-\beta U(x,t)}\frac{\partial}{\partial x} \left( e^{\beta U(x,t)} G   \right)\right],
\end{equation}
with initial condition
$G(x,t^\prime | x^\prime,t^\prime) = \delta(x-x^\prime)$.  
The
connection between $P$ and $G$ arises from by noting that $G(x,t|0,0)$
can be decomposed into two parts: (i) a contribution $P(x,t)$ from
those trajectories that never reach $x^{\ddagger}_0$ at any time prior to $t$; (ii)
a contribution from those trajectories that reach $x^{\ddagger}_0$ for the first
time at some $t^\prime \le t$, and then diffuse from $x^{\ddagger}_0$ to $x$ in the
time $t-t^\prime$.  The distribution of first passage times to $x^{\ddagger}_0$ is
just $-d\Sigma_r(t)/dt$, and the probability of getting from
$x^{\ddagger}_0$ to $x$ is $G(x,t|x^{\ddagger}_0,t^\prime)$.  Putting everything together, we
have
\begin{equation}\label{si3}
G(x,t|0,0) = P(x,t) - \int_0^t dt^\prime G(x,t|x^{\ddagger}_0,t^\prime) \frac{d\Sigma_r(t^\prime)}{dt^\prime}.
\end{equation}
Solving Eq.~\eqref{si3} for
$P(x,t)$, and then integrating $x$ from $-\infty$ to $x^{\ddagger}_0$, gives the
following integral equation for $\Sigma_r(t)$~\cite{Hu2010},
\begin{equation}\label{si4}
\Sigma_r(t) = \int_{-\infty}^{x^{\ddagger}_0} dx\, P(x,t) =  \int_{-\infty}^{x^{\ddagger}_0} dx\, G(x,t|0,0) + \int_0^t dt^\prime \frac{d\Sigma_r(t^\prime)}{dt^\prime} \int_{-\infty}^{x^{\ddagger}_0} dx\,G(x,t|x^{\ddagger}_0,t^\prime).
\end{equation}
To make further progress, we note that the solution to Eq.~\eqref{si2} for our choice of $U(x,t)$  is
\begin{equation}\label{si5}
\begin{split}
&G(x,t|x^\prime,t^\prime) = \frac{1}{\sqrt{2\pi\sigma(t-t^\prime)}} \exp\left(-\frac{\left(x-\mu(x^\prime,t-t^\prime)\right)^2}{2 \sigma(t-t^\prime)}\right),\\
&\sigma(t) \equiv \frac{1-e^{-2\beta D \omega_0 t}}{\beta \omega_0}, \quad
\mu(x,t) \equiv \frac{r(\beta D \omega_0 t -1) + e^{-\beta D \omega_0 t}(r + \beta D \omega_0^2 x)}{\beta D \omega_0^2},
\end{split}
\end{equation}
for $t \ge t^\prime$.  This describes a Gaussian function with
time-dependent mean $\mu$ and variance $\sigma$.  In the limit
$r\to\infty$ the mean $\mu$ rapidly increases with $t$, and the
character of the dynamics becomes more ballistic than diffusive.  As a
result the contribution (ii) described above, from those trajectories
that diffuse backward from $x^{\ddagger}_0$ to some $x \le x^{\ddagger}_0$, becomes
negligible.  Thus $G(x,t|0,0) \approx P(x,t)$ for $r\to\infty$, and we
can approximate Eq.~\eqref{si4} as
\begin{equation}\label{si6}
\begin{split}
\Sigma_r(t) &\approx  \int_{-\infty}^{x^{\ddagger}_0} dx\, G(x,t|0,0)\\
&= \frac{1}{\sqrt{2\pi\sigma(t)}} \int_{-\infty}^{x^{\ddagger}_0} dx\, \exp\left(-\frac{\left(x-\mu(0,t)\right)^2}{2 \sigma(t)}\right)\\
&= \frac{1}{2}\left(1+\text{erf}\left[\frac{x^{\ddagger}_0-\mu(0,t)}{\sqrt{2 \sigma(t)}} \right] \right)
\end{split}
\end{equation} 
Plugging in the values of $\mu(0,t)$ and $\sigma(t)$ from
Eq.~\eqref{si5}, and making the change of variables $f = r t$, gives
the approximate expression for $\Sigma_r(f)$ in Eq.~(9) of the main text.

\section{Sensitivity of the heterogeneity analysis to experimental artifacts}\label{exp}

All five sets of experimental data that exhibited heterogeneity in
Fig.~6 of the main text were collected using AFM pulling experiments.
(In contrast three of the collapsed data sets in Fig.~5 were from
optical tweezer studies, while the other two used AFM.)  Thus it is
important to check whether any aspects of the AFM experimental
apparatus or procedure could influence the analysis of heterogeneity.
We will consider three separate issues: cantilever force resolution
and drag, noise due to finite sampling of the rupture distributions,
and apparatus response time.  We will focus on the AFM case, since
this is most relevant to the existing data, but the discussion can
easily be generalized to optical tweezers.

\vspace{1em}

\noindent i) {\it Cantilever force resolution and drag:} AFM
cantilevers typically have spring constants
$\omega_c \sim {\cal O}(10\:\text{pN/nm})$.  Thermal fluctuations of
the cantilever limit the resolution at which forces can be measured to
$\delta f \sim \sqrt{\omega_c k_B T}$, where for example
$\delta f \sim 6$ pN when $\omega_c = 10$ pN/nm~\cite{Neuman2008}.
(Low-pass filtering of the data can in principle improve the force
resolution~\cite{Viani1999,Neuman2008}, but is not necessarily helpful
for experiments involving steep force ramps, where maximum temporal
resolution is necessary to pinpoint the rupture force.)  The
cantilever is also subject to viscous drag, with friction coefficient
$\gamma(h)$ that in general depends on the geometry of the cantilever
and its height $h$ from the surface.  Experimental measurements of
this drag are often fit well by a phenomenological scaled spherical
model, $\gamma(h) = 6 \pi \eta a_\text{eff}^2/(h+ h_\text{eff})$,
where $\eta = 0.89$ mPa$\cdot$s is the viscosity of the surrounding
water at room temperature, and $a_\text{eff}$ and $h_\text{eff}$ are
parameters with dimensions of length~\cite{Alcaraz2002,Janovjak2005}.
We will choose typical experimental values of $a_\text{eff} = 25$
$\mu$m and $h_\text{eff} = 5$ $\mu$m, and assume that the rupture
measurements are all conducted at $h \ll h_\text{eff}$, so in our
analysis the drag coefficient is approximately constant, with a value
$\gamma \approx 2$ pN$\cdot$s/$\mu$m (which matches the measured drag
coefficient in Ref.~\cite{Wojcikiewicz2006}).

An unloaded (post-rupture) cantilever moving at fast pulling speeds of
$v > 1$ $\mu$m/s (or ramp rates $r > 10^4$ pN/s for $\omega_c = 10$
pN/nm) away from the surface will feel drag forces
$f_\text{drag} = \gamma v > 2$ pN.  Since in experiments the magnitude
of the rupture force is defined as the difference in the pre-rupture
and post-rupture force levels, the drag creates a velocity-dependent
artifact.  The magnitude of the error in the measured rupture force
depends also on velocity of the cantilever tip pre-rupture, and hence
the stiffness of the sample: the softer the sample, the smaller the
velocity difference of the tip pre- and post-rupture, and the smaller
the error~\cite{Alcaraz2002,Janovjak2005,Liu2010}.  However in typical
biomolecule rupture experiments the sample at the point of rupture is
maximally extended, with large stiffness, and the tip velocity is much
slower than the pulling velocity.  In the limit where tip velocity
immediately pre-rupture is zero, the error reaches its maximal value:
the measured rupture force is approximately $\gamma v$ smaller than
the actual one due to the drag offset post-rupture.  This
underestimation has been observed in fast AFM pulling experiments on
the I27 domain of titin~\cite{Janovjak2005,Liu2010}.

To see the effects of cantilever artifacts on the heterogeneity
analysis, we compared two different numerical approaches for the FBL
model: a) The approach described in the main text and SI Sec.~\ref{sim}
(with results in Fig.~2 of the main text).  The simulations have an
idealized force ramp $f(t) = r t$ at fixed $r$ with no cantilever
artifacts.  The rupture force in a simulation trajectory is just
recorded as $r t_\text{rup}$, where $t_\text{rup}$ is the time of
rupture.  b) An analogous approach, using the Hamiltonian
$U(x,t) = (1/2) \omega_0 x^2 + (1/2) \omega_c (x_c(t) - x)^2$.  Here
$x_c(t)$ mimics the experimentally-controlled position of the clamped
end of the cantilever, with the tip end-point assumed to be at $x$,
and hence subject to thermal fluctuations. The cantilever stiffness is
set to $\omega_c = 10$ pN/nm.  To achieve an average ramp rate of $r$,
the position $x_c(t) = v t$, with the velocity chosen to be
$v = r/\omega_c$.  The rupture force for a simulation trajectory is
recorded as $\omega_c (v t_\text{rup} - x_\text{rup}) - \gamma v$,
where $x_\text{rup}$ is the value of $x$ at rupture, and the
$\gamma v$ offset reflects the worse case scenario for drag-induced
error.

\begin{figure}
\includegraphics[width=\textwidth]{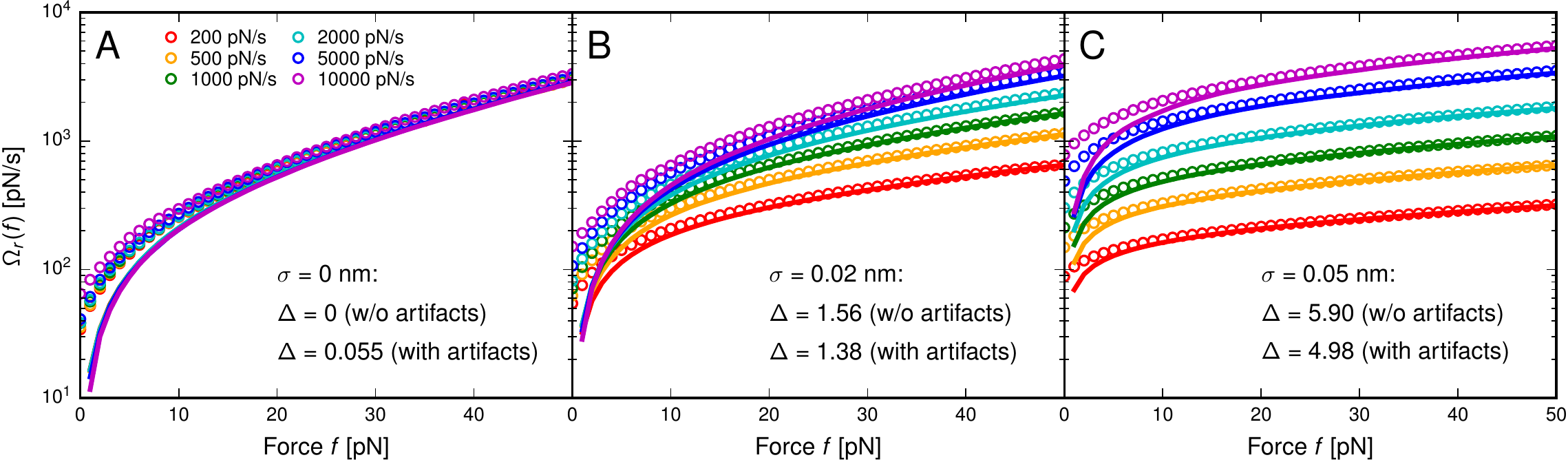}
\caption{Analysis of the FBL heterogeneous model system, using the two
  different numerical approaches described in SI Sec.~\ref{exp}.i.  The
  parameters are the same as in Fig.~2 of the main text, and we show
  simulation results for three cases with $k_i = 0$: A) $\sigma=0$ nm;
  B) $\sigma = 0.02$ nm; C) $\sigma=0.05$ nm.  The first numerical
  approach (solid curves) does not include cantilever artifacts, while
  the second (circles) does.  The best-fit values of the heterogeneity
  parameter $\Delta$ from both approaches are listed in each
  panel.}\label{noise}
\end{figure}

Fig.~\ref{noise} shows the numerical results for $\Omega_r(f)$ in the
quenched disorder limit ($k_i = 0$) using the two approaches, with
solid curves representing case a) and circles case b).  Panels A
through C correspond to different levels of disorder: $\sigma = 0$,
$0.02$, $0.05$ nm, and in each panel the range of ramp rates is
$r = 200 - 10000$ pN/s, comparable to the rates used in the
experimental observations of heterogeneity (main text Fig.~6).  The
two numerical approaches converge as $f$ increases, but show clear
discrepancies in the low force regime ($\lesssim 10$ pN), a
consequence of the cantilever artifacts.  Despite these artifacts, the
extracted heterogeneity parameters $\Delta$ from the two approaches
are similar.  In panel A ($\sigma = 0$ nm, no heterogeneity), we are
close to total collapse even in the presence of artifacts, with a
$\Delta$ value near zero.  In panels B and C ($\sigma = 0.02$ and
$0.05$ nm) the $\Delta$ values of the second approach differ from the
first one by less than $16\%$ due to the artifacts.  In all these
cases there are sufficient data points at larger forces (f
$\gtrsim 10$ pN) to mitigate the cantilever effects, and give a robust
estimation of $\Delta$.  We note that all the experimental data sets
in Fig.~6 of the main text entirely fall in this larger force regime,
and thus should yield reliable values for $\Delta$, even without
correcting for drag artifacts.  (Though in at least one of the
studies, corresponding to panel A of Fig.~6, the researchers
explicitly corrected for cantilever drag in measurements at pulling
speeds of $v > 1$ $\mu$m/s~\cite{Wojcikiewicz2006}.)

\vspace{1em}

\noindent ii) {\it Finite sampling:} Since $\Omega_r(f)$ depends on
the survival probability distribution $\Sigma_r(f)$, the analysis of
heterogeneity is sensitive to sampling noise in this distribution.  In
typical experiments the number of rupture events recorded at each $r$
is $\sim {\cal O}(10^2)$, and thus it is useful to determine the
uncertainty in the best-fit values of $\Delta$ due to the finite
sampling of the distribution.  To do this, we investigated every
heterogeneous experimental system in Fig.~6 of the main text, and
carried out the following procedure: the minimal model best-fit
theoretical result for $\Omega_r(f)$ was used to determine an
analytical form for the survival probability distribution
$\Sigma_r(f) = \exp(-\Omega_r(f)/r)$ at each experimental value of
$r$.  We then generated 1000 synthetic experimental data sets, drawing
$N_\text{ev}$ values of the rupture force $f$ from the cumulative
distribution $1-\Sigma_r(f)$ at every $r$ through inverse transform
sampling.  The value of $N_\text{ev}$ is listed for each experimental
system in Table~S1, and is based on the number of rupture
events per pulling speed measured in that particular study.  For each
of the 1000 synthetic data sets, the best-fit value of $\Delta$ was
extracted, and from the resulting distribution of $\Delta$ values we
calculated 95\% confidence intervals (also listed in
Table~S1).  The confidence intervals all lie within roughly
30\% of the original best-fit value of $\Delta$ in each system.  This
again reinforces the robustness of the experimental $\Delta$ values
determined in the main text.

\begin{table}
\bgroup
\def\arraystretch{1.5}
\begin{minipage}{\textwidth}
\centering
\setcounter{mpfootnote}{\value{footnote}}
\renewcommand{\thempfootnote}{\arabic{mpfootnote}}
\begin{tabular}{|l|c|c|c|}
\hline
System & $N_\text{ev}$ & $\Delta$ & 95\% CI for $\Delta$\\
\hline
ICAM1-LFA1~\cite{Wojcikiewicz2006} & 200\footnotemark[1]&1.5&$1.0-1.7$\\
Ran-imp$\beta$~\cite{Nevo2004} &375\footnotemark[2]& 2.6&$1.8-2.8$\\
DNA oligomer~\cite{Strunz1999} &300\footnotemark[2]&5.5&$4.2-6.3$\\
DNA-expG~\cite{Raible2006BJ} &200\footnotemark[1]&7.4&$5.5-8.0$\\
 RNA-{\it At}GRP8~\cite{Fuhrmann2009} &225\footnotemark[2]&13.3&$9.2-15.2$\\
\hline
\end{tabular}
%\vspace{-0.75\skip\footins}
\footnotetext[1]{In cases where the number of recorded rupture events is not specified in the study, we set $N_\text{ev} = 200$, a typical value.}
\footnotetext[2]{In cases where a range of $N_\text{ev}$ was reported, we chose the mean value of the range.}

\renewcommand{\footnoterule}{}
\end{minipage}
\egroup
\caption{Analysis of finite sampling effects on the determination of $\Delta$ from the experimental data in Fig.~6 of the main text.  $N_\text{ev}$ is the number of rupture events at each ramp rate in the experimental study, $\Delta$ is the best-fit theoretical value for the heterogeneity parameter, and the last column shows the 95\% confidence interval (CI) for $\Delta$.  The CI's are based on results from 1000 synthetic data sets, generated as described in SI Sec.~5.ii.}\label{tab1}
\end{table}

\vspace{1em}

iii) {\it Apparatus response time:} To more accurately describe the
experimental dynamics, the equilibration rate $k_\text{eq}$ should
reflect the overall relaxation time of the biomolecule plus apparatus
(i.e. AFM cantilever).  Depending on the details of the biological
system, either the biomolecule or apparatus might be rate-limiting in
determining $k_\text{eq}$.  If for example the apparatus response is
rate-limiting, and it makes $k_\text{eq}$ small enough that either
$\bar{k}(r) > k_\text{eq}$ or $k_c(r) > k_\text{eq}$, we would violate
the adiabatic condition.  The experimental consequences of this would
be similar to the largest ramp rates shown in Fig.~4 of the main text,
where we examined the non-adiabatic limit. There would be no collapse
in the $\Omega_r(f)$ curves, but the qualitative behavior would be
very different from the heterogeneous case: the non-adiabatic
$\Omega_r(f)$ curves grow further and further apart as $r$ is
increased.  For a given $f$ the non-adiabatic $\Omega_r(f)$ curve
decreases with increasing $r$, the opposite of the behavior in the
heterogeneous case.  However we see no evidence of non-adiabatic
behavior in any of the experimental data sets in either Fig.~5
(non-heterogeneous) or Fig.~6 (heterogeneous) of the main text.  This
indicates that the respective instrument relaxation rates must all be
larger than the lower bounds on $k_\text{eq}$ shown in Fig.~5F and 6F.
In the AFM case, the relaxation rate of a cantilever with stiffness
$\omega_c$ and friction coefficient $\gamma$ is $\omega_c/\gamma$.
Using typical values of $\omega_c = 10$ pN/nm and $\gamma = 2$
pN$\cdot$s/$\mu$m, we get $\omega_c/\gamma = 5000$ s$^{-1}$, which is
indeed larger than the lower bounds depicted in the figures.

\section{Relative likelihood analysis of heterogeneous vs. pure model fitting for experimental data}\label{like}

As further validation of our heterogeneity analysis, we compared the
likelihood ${\cal P}({\cal D}|{\cal M})$ of obtaining the experimental
data ${\cal D}$ (the histograms for the rupture force distributions
$p_r(f)$) given two choices of theoretical model ${\cal M}$:

\begin{enumerate}
\item {\it Heterogeneous model} ${\cal M}_\text{het}$: from Eqs. (6)
  and (7) in the main text, this model yields an analytical form for
  $\Omega_r(f)$ based on three parameters: $\Delta$, $k_0$, and
  $x^\ddagger$.  The predicted rupture force distribution $p_r(f)$ is
  given by:
\begin{equation}\label{l1}
\begin{split}
p_r(f) = -\frac{d\Sigma_r(f)}{df} &= -\frac{d}{df} e^{-\Omega_r(f)/r}= \frac{k_0 e^{\beta f x^\ddagger}}{r}\left(1+\frac{\Delta k_0 (e^{\beta f x^\ddagger}-1)}{\beta r x^\ddagger} \right)^{-\frac{\Delta+1}{\Delta}}
\end{split}
\end{equation}

\item {\it Pure model} ${\cal M}_\text{pure}$: this model assumes that
  we are pulling adiabatically on a system with a single functional
  state, with rupture described by the DHS~\cite{Dudko2006} rate in
  Eq.~\eqref{t3}, which is the most widely used theoretical fitting
  form in the pure case.  The predicted rupture force distribution
  $p_r(f)$ for this model is:
\begin{equation}\label{l2}
\begin{split}
p_r(f) &= \frac{k_0 \left(1 - \frac{\nu f x^\ddagger}{G^\ddagger}\right)^{-\frac{\nu-1}{\nu}}}{r}\exp\Biggl[\beta G^\ddagger\Biggl(1 -\left(1 - \frac{\nu f x^\ddagger}{G^\ddagger}\right)^{1/\nu} \Biggr)\\
&\qquad\qquad\qquad\qquad + \frac{k_0}{\beta r x^\ddagger}\Biggl(1-e^{\beta G^\ddagger\left(1 -\left(1 - \nu f x^\ddagger/G^\ddagger\right)^{1/\nu} \right) }\Biggr)\Biggr]
\end{split}
\end{equation}
Setting $\nu=2/3$ (the other choice $\nu=1/2$ gives similar results)
this model then depends on three fitting parameters: $k_0$,
$x^\ddagger$, and $G^\ddagger$.  The DHS model reduces to the pure
Bell theory when $G^\ddagger \to \infty$.  Since the $\Delta \to 0$
limit of the heterogeneous model also yields the pure Bell theory,
Eq.~\eqref{l1} when $\Delta \to 0$ and Eq.~\eqref{l2} when
$G^\ddagger \to \infty$ are equivalent.  Away from those limits, the
two models give different results for $p_r(f)$.

\end{enumerate}

\begin{figure}
\includegraphics[width=0.5\textwidth]{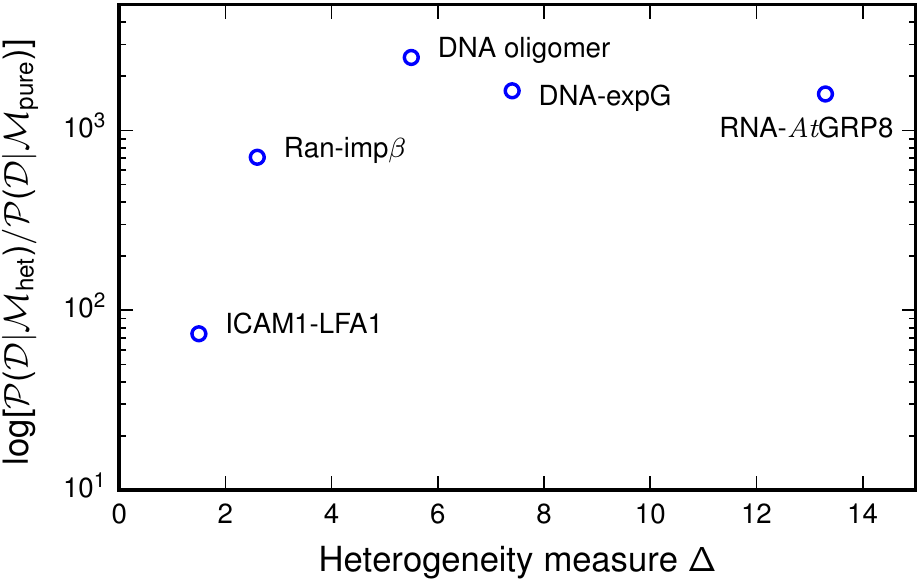}
  \caption{The logarithm of the relative likelihood,
    $\log[{\cal P}({\cal D}|{\cal M}_\text{het})/{\cal P}({\cal D}|{\cal
      M}_\text{pure})]$ of the heterogeneous model versus the pure
    model for the experimental data from five studies: ICAM1-LFA1
    ~\cite{Wojcikiewicz2006}, Ran-imp$\beta$~\cite{Nevo2004}, DNA
    oligomer~\cite{Strunz1999}, DNA-expG~\cite{Raible2006BJ}, RNA-{\it
      At}GRP8~\cite{Fuhrmann2009}.  The log-relative-likelihood is
    plotted on the vertical axis, while the horizontal axis shows the
    corresponding results for the heterogeneity parameter
    $\Delta$.}\label{lrat}
\end{figure}

For each experimental system, there are measurements from
$N^\text{load}$ different loading rates $r_\rho$,
$\rho=1,\ldots,N^\text{load}$.  The data at each loading rate are
given as a set of $N^\text{hist}_{\rho}$ histogram counts
$\{f_{\rho,i}, N_{\rho,i}\}$, $i=1,\ldots,N^{\text{hist}}_{\rho}$,
where $f_{\rho,i}$ is the force at which the $i$th bin is centered,
and $N_{\rho,i}$ is the number of experimental trajectories which
ended in rupture at a force $f$ that fell within the bin range
$f_{\rho,i} - w/2 < f < f_{\rho,i}+w/2$.  Here $w$ is the width of the bin.  For any
loading rate the total number of events is taken to be constant,
$N_\text{ev} = \sum_i N_{\rho,i}$, with the values of $N_\text{ev}$
for each experimental system we analyzed listed in Table~S1.  The
assumption of constant $N_\text{ev}$ is due to the fact that most
studies did not explicitly list the individual values of
$\sum_i N_{\rho,i}$ for each $\rho$, but instead gave a typical range.
For a given model ${\cal M}$ and its corresponding set of parameter
values, the probability of observing an experimental rupture outcome
that falls within the $f_{\rho,i}$ bin is:
\begin{equation}\label{l3}
{\cal P}_{\rho,i}({\cal M}) = \int_{f_{\rho,i}-w/2}^{f_{\rho,i}+w/2} df\, p_r(f)
\end{equation}
with $p_r(f)$ given by either Eq.~\eqref{l1} or \eqref{l2} depending
on ${\cal M}$.  The overall likelihood of all the experimental
outcomes for a system is:
\begin{equation}\label{l4}
{\cal P}({\cal D}|{\cal M}) = \prod_{\rho=1}^{N^\text{load}} \prod_{i=1}^{N^\text{hist}_\rho} [{\cal P}_{\rho,i}({\cal M})]^{N_{\rho,i}}
\end{equation}
The relative likelihood
${\cal P}({\cal D}|{\cal M}_\text{het})/{\cal P}({\cal D}|{\cal
  M_\text{pure}})$ is a measure of how much more likely it is that the
heterogeneous model describes the experimental data compared to the
pure model.  Note that both models depend on the same number of
parameters.  The value of ${\cal P}({\cal D}|{\cal M}_\text{het})$ for
each experimental system with nonzero $\Delta$ is calculated using the
parameters listed in Fig.~6 of the main text.  For the pure model, we
found the parameter set $k_0$, $x^\ddagger$, and $G^\ddagger$ that
maximizes ${\cal P}({\cal D}|{\cal M}_\text{pure})$ and used that
maximum likelihood value for the comparison.  In Fig.~\ref{lrat} we
plot the logarithm of the relative likelihood,
$\log[{\cal P}({\cal D}|{\cal M}_\text{het})/{\cal P}({\cal D}|{\cal
  M_\text{pure}})]$ on the vertical axis for the experimental systems,
versus the corresponding value of the heterogeneity parameter $\Delta$
on the horizontal axis.  All the relative likelihoods overwhelmingly
favor the heterogeneous model.  Even the smallest relative likelihood,
which coincides with the smallest $\Delta$ value (ICAM1-LFA1 with
$\Delta = 1.5$) is still highly favorable for heterogeneity, with a
ratio
${\cal P}({\cal D}|{\cal M}_\text{het})/{\cal P}({\cal D}|{\cal
  M_\text{pure}}) \approx 10^{32}$.  Thus we can conclude that for the
systems identified as heterogeneous by their $\Delta$ values
(corresponding to non-collapse of the $\Omega_r(f)$ curves), the best
available pure model is an extremely unlikely alternative description.
The collective data for each system, representing hundreds of
experimental trials, unambiguously points to heterogeneity.

\end{document}